\documentclass[%
 reprint,
amsmath,amssymb,
prb,
showkeys
]{revtex4-2}

\usepackage[utf8]{inputenc}
\usepackage[margin=1in]{geometry}
\usepackage{blindtext}
\usepackage[utf8]{inputenc}
\usepackage{braket}
\usepackage{amsmath} 
\usepackage{graphicx}
\usepackage{breqn}
\usepackage{amsmath}
\usepackage{amssymb}
\usepackage{amsthm}
\usepackage{linop}  
\usepackage{mathtools} 
\usepackage{comment}
\usepackage{placeins}
\usepackage{natbib} 
\usepackage{lipsum}
\usepackage{xcolor}
\usepackage{hyperref}

\newcommand{\ii}{i}
\renewcommand{\refeq}[1]{Eq.~(\ref{eq:#1})}   
\newcommand{\reffig}[1]{Figure~\ref{fig:#1}}
\newcommand{\Tr}{\text{Tr}}

\makeatletter
\let\cat@comma@active\@empty
\makeatother

\begin{document}



\title{The Lieb-Robinson correlation function for long disordered transverse-field Ising chains}  

\author{Brendan J. Mahoney}
\affiliation{ 
Department of Electrical Engineering\\
University of Notre Dame\\
Notre Dame, IN~46556, USA
}%
\author{Craig S. Lent}
\affiliation{ 
Department of Electrical Engineering\\
Department of Physics and Astronomy\\
University of Notre Dame\\
Notre Dame, IN~46556, USA
}%

\date{\today}

\begin{abstract}
The Lieb--Robinson correlation function can be used to characterize the propagation of quantum information in  Ising chains.  Considerable work has been done to establish \emph{bounds} on this correlation function in various circumstances. To actually calculate the \emph{value} of the correlation function directly typically requires a state space which grows exponentially with system size, and so is intractable for all but relatively small systems. We employ a recently-developed method that enables direct calculation of the  Lieb--Robinson correlation function and which scales linearly with system size. This enables the computation for systems with more than a thousand  qubits, revealing the propagation of quantum information down the chain. We extend this technique to the problem of  Ising chains with randomly disordered coupling strengths. Increasing disorder causes localization of the quantum correlations and halts propagation of quantum information
down the chain. We see light-cone isocontours therefore becoming vertical. The decay of correlations down the chain is characterized by a correlation length which in the paramagnetic phase increases quadratically with mean coupling strength and inverse quadratically with disorder level. The correlation length peaks  at the phase transition. We map our general expression for the correlation function in the Ising system onto the probability density in the single particle tight-binding Su-Schrieffer–Heeger model for a specific initial condition. 

\end{abstract}

\keywords{Lieb--Robinson, quantum correlations, entanglement, Ising model, Anderson localization, many-body localization}

\maketitle

\newcommand\figwidth{1.0\columnwidth}


\section{Introduction}

The spread of information and correlations in quantum many-body systems is constrained in a fundamental way, even in the absence of relativistic causality. Lieb and Robinson showed that for a broad class of lattice Hamiltonians with local interactions, the growth of operator support is effectively limited by a finite velocity~\cite{Lieb1972}. 

Their analysis considers local operators $\hat{A}_m$ and $\hat{B}_k$ acting on distinct subsystems $k$ and $m$, with the time evolution of $\hat{A}_m$ given in the Heisenberg picture by
\begin{equation}
\hat{A}_m(t) = e^{i \hat{H} t/\hbar} \, \hat{A}_m(0) \, e^{-i \hat{H} t/\hbar}.
\label{eq:HeisenbergTimeDependenceIntro}
\end{equation}
The growth of correlations between the two subsystems can be monitored through the norm of the commutator
\begin{equation}
    C_{A,B}(t) = \bigl\| [\hat{A}_m(t), \hat{B}_k] \bigr\|.
    \label{eq:LRcommutatorIntro}
\end{equation}
\noindent If $k\ne m$, $C$ is zero at $t=0$ because the operators act on distinct subsystems so the two operators commute. As time passes, $\hat{A}_m(t)$ spreads under \refeq{HeisenbergTimeDependenceIntro} to include components with operators on subsystems other than $m$. When   $\hat{A}_m(t)$ has significant components which act on subsystem $k$, the commutator can become non-zero, signaling the emergence of specifically quantum correlations between $m$ and $k$.

Under general locality conditions, this quantity obeys the Lieb--Robinson bound
\begin{equation}
    C_{A,B}(t) \leq K \, e^{-\mu \bigl(d_{k,m}-v_{\text{\tiny LR}} t\bigr)},
    \label{eq:BasicLRboundIntro}
\end{equation}
where $d_{k,m}$ is a measure of the distance between the subsystems, $K$ and $\mu$ are system-dependent constants and 
$v_{\text{\tiny LR}}$ is the Lieb-Robinson velocity.  Outside the effective light cone defined by this velocity, the commutator is
exponentially small. This result has far-reaching implications, providing
rigorous bounds on the locality of non-relativistic quantum dynamics and
establishing that correlations and entanglement can only propagate
within a finite causal region set by the Lieb--Robinson velocity~\cite{Hastings2006,Nachtergaele2006}.

Considerable work has since refined and extended this framework. For short-range interacting Hamiltonians, tighter bounds have been obtained using a commutative graph construction~\cite{Wang2020}, with subsequent improvements in related approaches~\cite{Braida2023,Hinrichs2025}. Other directions have focused on systems with long-range couplings, where power-law interactions give rise to modified, nonlinear light cones~\cite{Sweke2019,Luitz2020,Cevolani2016,Tran2020,Schneider2021}. These developments highlight the range of settings in which Lieb--Robinson techniques are now applied \cite{Bru2016, Chen2023a, Lucas2021b}.

Disorder represents a distinct and particularly rich direction. Anderson's work established that in one and two dimensions, arbitrarily weak disorder localizes all single-particle states, while in three dimensions a localization transition occurs at finite disorder~\cite{Anderson1958,Frohlich1983}. This localization arises from quantum interference between closed-loop paths that enhances return probabilities and suppresses transport. In interacting spin systems, the impact of disorder was explored early on in studies of the transverse-field Ising model~\cite{McCoy1968,Pfeuty1979}. Building on Griffiths' original work on random ferromagnets~\cite{Griffiths1969}, later analyses revealed how rare, strongly ordered regions can appear inside the disordered phase, producing unusual scaling behavior and strongly affecting the dynamics of the system~\cite{Igloi2005,Kisker1998,Hoyos2012}.

When disorder is introduced, the standard linear light-cone structure implied by
the Lieb--Robinson bound can be substantially modified. 
General analyses have revealed the emergence of logarithmic light cones in
several disordered models, including the XY chain and related spin systems
~\cite{Burrell2007,Friesdorf2015,Zeng2025,Baldwin2023,Baldwin2025, Gebert2016,Damanik2016}. In many of these cases, the commutator obeys a bound of the form
\begin{equation}
C_{A,B}(t) \;\leq\; K |t|^\alpha \, e^{-\frac{d_{k,m}}{\xi}}
\;=\; K \, e^{\alpha\log(|t|) - \frac{d_{k,m}}{\xi}} ,
\label{eq:LiebRobinsonboundDisorderedIntro}
\end{equation}
where $K$ and $\xi$ are system-dependent constants that set the overall
scale and spatial decay length of correlations, respectively. Here the time
dependence enters only as a power-law rather than exponentially, resulting in an
effective light cone that expands logarithmically in time. This behavior reflects
the  suppression of operator and information spreading created by
disorder, and provides a  characterization of slow entanglement growth and constrained locality in disordered quantum phases.
For the XY model with disorder,  the Lieb--Robinson correlation function has been shown to be bounded under some conditions by a time-independent envelope \cite{Hamza2012, Stolz2017}. 

It is usual to study  either \emph{bounds} on $C$, or to calculate the actual values of $C$ for relatively small systems. The reason for this is the familiar exponential explosion in the size of the state space for a  multi-qubit or spin system. 

We focus here on the simplest dynamical system, a one-dimensional transverse-field Ising model (TFIM), with the aim of calculating the value of the Lieb--Robinson (LR) correlation function for large systems and exploring the effects of disorder. To this end, we extend our recently developed method \cite{Lent2024} for computing the LR correlation function on regular TFIM  chains. 

The simplicity of the TFIM model does not mean it is without real-world applications \cite{Igloi2026}. It is a frequently applicable model for qubit arrays. Systems with thousands of qubits are now becoming available, making it possible to realize and study large many-body systems in controlled settings. Examples include entanglement generation in gate-defined quantum dots~\cite{Steinacker2025}, coherent evolution in neutral-atom arrays~\cite{Chiu2025}, and quantum annealing experiments on large TFIM systems~\cite{King2025}. These developments make efficient theoretical methods for analyzing LR correlations in large systems increasingly important.

The present approach for studying the propagation of quantum correlations should be distinguished from state-based methods. A common technique is to solve the time-dependent Schr\"odinger equation for a state that is the ground state of an initial Hamiltonian $H_0$. At $t=0$, the Hamiltonian is switched to a new Hamiltonian $H$, and the time evolution of the state is calculated, using various exact or approximate means. Such ``quench'' calculations have the great advantage of yielding all possible information in the quantum state over time. The disadvantages are that the results depend on the initial state and, as usual, are limited by the feasibility of representing the state space adequately.  The LR correlation function is state-independent, and so depends only on the system Hamiltonian. It is also a two-time correlation function---the correlation is between one qubit at $t=0$ and another at some later time $t$. 
Finally, the time development of the LR correlation function is not a low-energy phenomenon; the full energy spectrum participates.

Section \ref{sec:PauliWalk} briefly reviews the transverse-field Ising model and the operator Pauli walk method for calculating the LR correlation function. The application of this method  to the disordered chain is presented in Section \ref{sec:DisorderedIsing}.
Section \ref{sec:SSHcomparison} discusses  mapping our main results from the TFIM onto the Su-Schrieffer-Heeger model. 
In Appendix A, we  review Anderson localization in a single-particle 1D tight-binding chain, using  techniques and  visualizations we apply to the TFIM. This even-simpler model provides a useful point of comparison.  Appendix B provides an alternate formulation of the results of the operator Pauli walk method using Majorana operators.


\section{Operator Pauli walk method for the Ising chain \label{sec:PauliWalk}}

\subsection{Transverse field Ising model}

We consider a 1D chain of $N_q$ interacting qubits in states $\ket{0}$ or $\ket{1}$, or equivalently of spins $\ket{\downarrow}$ and $\ket{\uparrow}$ (we will use the description interchangeably). The system is described by the  transverse-field Ising model (TFIM) Hamiltonian with nearest-neighbor couplings: 
\begin{equation}
H \;=\; - J \sum_{k=1}^{N_q-1}  Z_k Z_{k+1} \;-\; \gamma \sum_{k=1}^{N_q} X_k ,
\label{eq:H_dimful}
\end{equation}
Here  $J$ denotes the coupling energy between qubits; we take $J\ge 0$. For a spin system, the  $X$ term can represent an applied transverse magnetic field. We interpret it here equivalently as representing the kinetic energy associated with the flipping dynamics of each qubit or spin. The presence of this term means that an isolated $\ket{1}$  (or $\ket{\uparrow}$) state is not a stationary state, but will precess in time.  A natural timescale, related to the Rabi oscillation time for an isolated qubit, is in terms of the quantity
\begin{equation}
\tau \;=\; \frac{\pi \hbar}{\gamma}.
\label{eq:tau_def}
\end{equation}

It is helpful then to measure  time in terms of $\tau$ and energy in terms of $\gamma$. We define the dimensionless Hamiltonian
\begin{equation}
H' \equiv H/\gamma
=- J' \sum_{k=1}^{N_q-1}  Z_k Z_{k+1} \;-\; \sum_{k=1}^{N_q} X_k.
\label{eq:H_dimless}
\end{equation}
where $J' \equiv J/\gamma$. 

\begin{figure}[tbp]
    \centering
    \includegraphics[width=\figwidth]{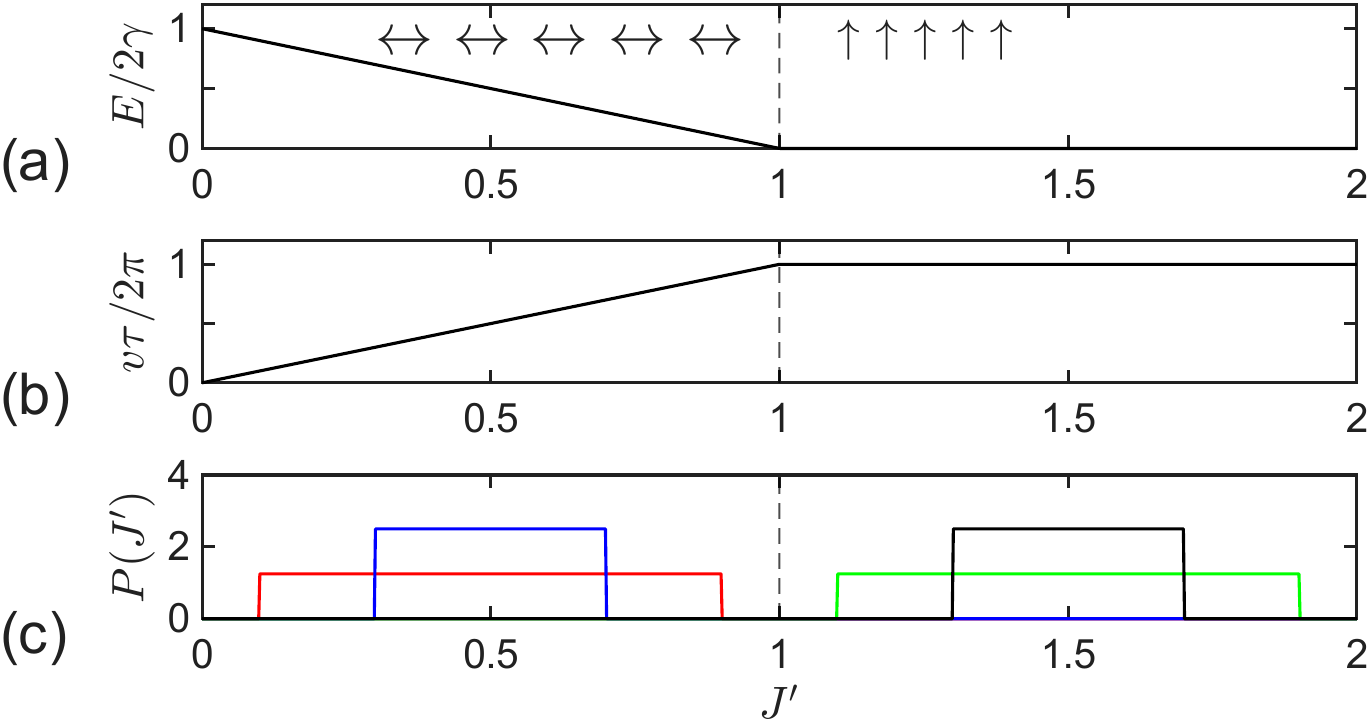}
    \caption{
    Two phases of the TFIM chain. 
    (a) Energies of the ground state and first excited state as a function of coupling $J'$. For $J'<1$, the system is in the paramagnetic state; for $J'>1$ it is in the ferromagnetic state.
    (b) Front velocity as a function of the coupling $J'$ given by \refeq{vFront}.
    (c) Probability distributions for disordered couplings as given by \refeq{UniformPDF_inh} with $J'_0=0.5$,  $\Delta J'=0.4$ (blue) and $\Delta J'=0.8$ (red);  $J'_0=1.5$,  $\Delta J'=0.4$ (black) $\Delta J'=0.8$ (green). 
    }
    \label{fig:multi_panel_ising}
\end{figure}

The TFIM has two quantum phases separated by a critical point at $J'=1$.
Figure \ref{fig:multi_panel_ising}(a) shows the ground state  and first excited state energies for the system in the 
thermodynamic limit. 

For $J'\le 1$ we say the system is in the \emph{paramagnetic} phase. When $J'=0$, the chain consists of noninteracting spins whose energy eigenstates are each eigenstates of $X_k$: the symmetric ground state $\ket{X^+}=(\ket{\uparrow} + \ket{\downarrow})/\sqrt{2}$ or antisymmetric excited state  $\ket{X^-}=(\ket{\uparrow} - \ket{\downarrow})/\sqrt{2}$. The ground state of the chain consists of all spins in the ground state. The first excited state of the system consists of a single site excited to the $\ket{X^-}$ state, which costs an energy $2\gamma$. As the coupling $J'$ is turned on, a single excitation on one site can move down the chain, delocalizing and lowering the energy.  In this regime, we can think of the system quasiparticles as being roughly these mobile excitations.

For $J'> 1$,  the system is in the \emph{ferromagnetic} phase. For $J>>\gamma$, the ground state of the whole chain consists of all  sites in $\ket{\uparrow}$ state or all in the $\ket{\downarrow}$ state. The $Z_2$ symmetry (flipping all the spins) produces a two-fold degenerate ground state. For finite $J'>1$, we can think of the quasiparticle excitations as domain walls in the chain: $\ket{\cdots \uparrow \uparrow \uparrow \downarrow \downarrow \downarrow \cdots}$. The motion of the domain wall is enabled by the nonzero kinetic energy term. 

For both the paramagnetic and ferromagnetic phases with $0 <J'<\infty$, actual system eigenstates are delocalized combinations of these  excitations, which mix the character of the simple quasiparticle descriptions above.

\begin{figure}[bt]
    \centering
    \includegraphics[width=\figwidth]{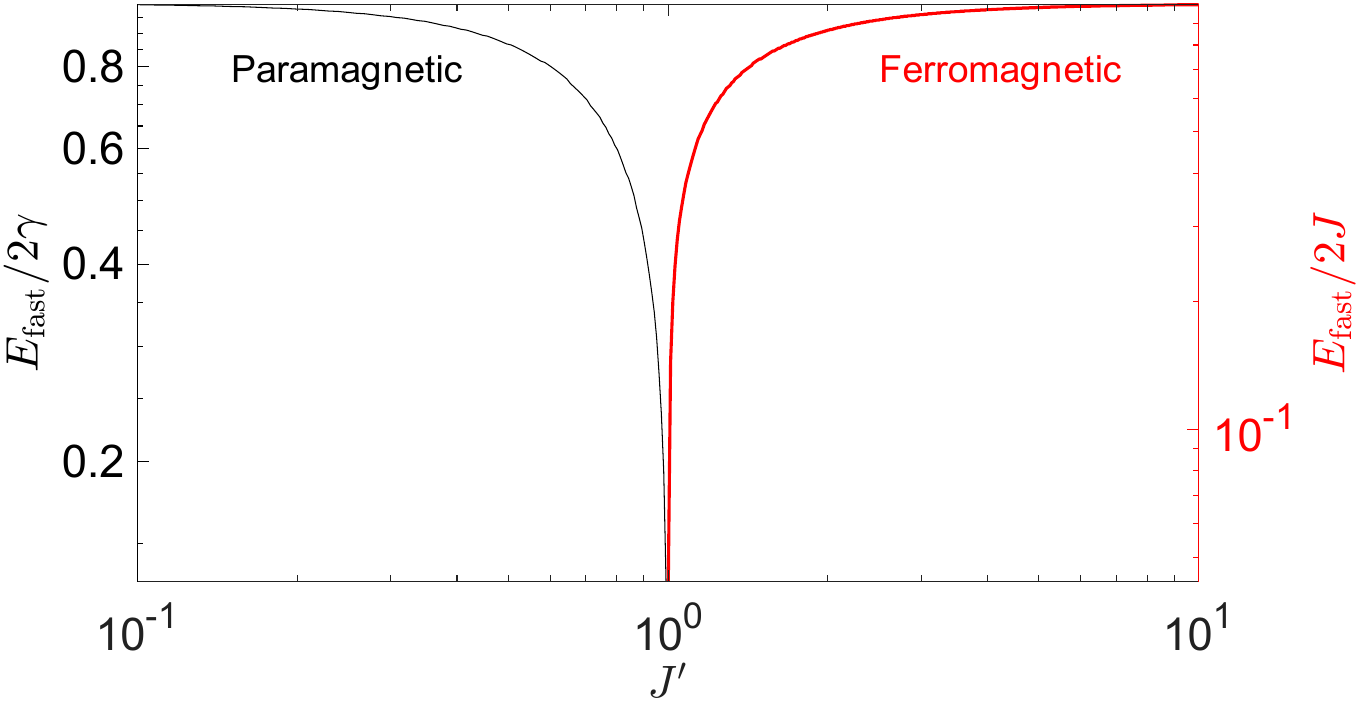}
    \caption{
    The energy of the fastest quasiparticle.
    }
    \label{fig:EfastJp}
\end{figure}

The TFIM can be mapped onto an equivalent fermion problem using the Jordan-Wigner transformation \cite{Lieb1961, Sachdev2011, Igloi2023}. A subsequent Bogoliubov transformation rotates the fermion creation and annihilation operators to yield a free fermion Hamiltonian with 
bulk  quasiparticle excitation spectrum
\begin{equation}
 E(q)= 2J\sqrt{g^2 + 1 - 2g\cos{q}} .
 \label{eq:epsQuasiparticle}
\end{equation}
Here $q$ is the wavenumber of the harmonic excitation and $g\equiv 1/J'$.
The energy eigenvalues for the system are all possible combinations of the quasiparticle excitation energies given by \refeq{epsQuasiparticle}.
The corresponding quasiparticle group velocities can be calculated via the usual expression $v(q)=(1/\hbar) \partial E/ \partial q$. The maximum value over $q$ of this  velocity is given by 
\begin{align}
    v\, \tau=
    \begin{cases}
       2\pi J'  &\quad    J'\le 1 \\
        2\pi         &\quad    J' > 1.
    \end{cases}
    \label{eq:vFront}
\end{align}
A derivation is given in \cite{Lent2024}.  Figure \ref{fig:multi_panel_ising}(b) shows this velocity as a function of $J'$. As we will see below, this velocity represents the velocity of the correlation front which propagates down the chain; we therefore refer to it as the front velocity. In the context of quantum chaos, it is also known as the butterfly velocity.

We can define $E_{\text{fast}}$ as the energy of the fastest quasiparticle, that is, the value of the  energy in \refeq{epsQuasiparticle} for which the group velocity equals the maximum value given by \refeq{vFront}.  \reffig{EfastJp} shows this quantity for both the paramagnetic and ferromagnetic phases. As suggested above $E_{\text{fast}}$  becomes $2\gamma$ in the extreme paramagnetic limit ($J'=0$) and $2J'$ in the extreme ferromagnetic limit ($J'>>1$). At the critical point, this energy goes to 0.

There is another velocity which characterizes the propagation of the first-order effect of correlations,  discussed in \cite{Mahoney2022, Lent2024}, but it reflects the initial turn-on of extremely small values of correlation and will not concern us here.

\subsection{The operator Pauli walk method for the Lieb--Robinson correlation function}
To characterize the dynamic spread of correlations, we study the
Lieb--Robinson correlation function
\begin{equation}
C_k(t) \;=\; \big\| [Z_1(t), Z_k] \big\|,
\label{eq:C_def_inh}
\end{equation}
which quantifies how correlations propagate as a function of time.  We use the normalized Frobenius norm
\begin{equation}
    \big\| \mathcal{O} \big\|= \sqrt{ \frac{1}{2^{N_q}} \Tr\left(\mathcal{O}^\dagger \mathcal{O}\right)   }.
    \label{eq:FroebNorm}
\end{equation}

The operator $Z_1(t)$ evolves under $\hat{H}$ according to
\begin{equation}
Z_{1}(t) = \sum_{n=0}^{\infty} \frac{1}{n!} \pi^n i^n 
\left(\frac{t}{\tau} \right)^n 
\left[\left(\hat{H}'\right)^{n},Z_1\right].
\label{eq:Z1Oft}
\end{equation}

From the definition in \refeq{C_def_inh}, we see that at $t=0$, $Z_1$ commutes with itself, and also with  $Z_k$,  which acts on a different site $k$, so $C_k(0)=0$. The time evolution of \refeq{Z1Oft} results in $Z_1(t)$ spreading to include components of Pauli operators on other sites.
It follows from the Pauli operator commutation relations (e.g. $[Z,X]=2iY$) that the maximum value of $C_k(t)$ is $2$.

\begin{figure}[!tbh]
    \centering
    \includegraphics[width=\columnwidth]{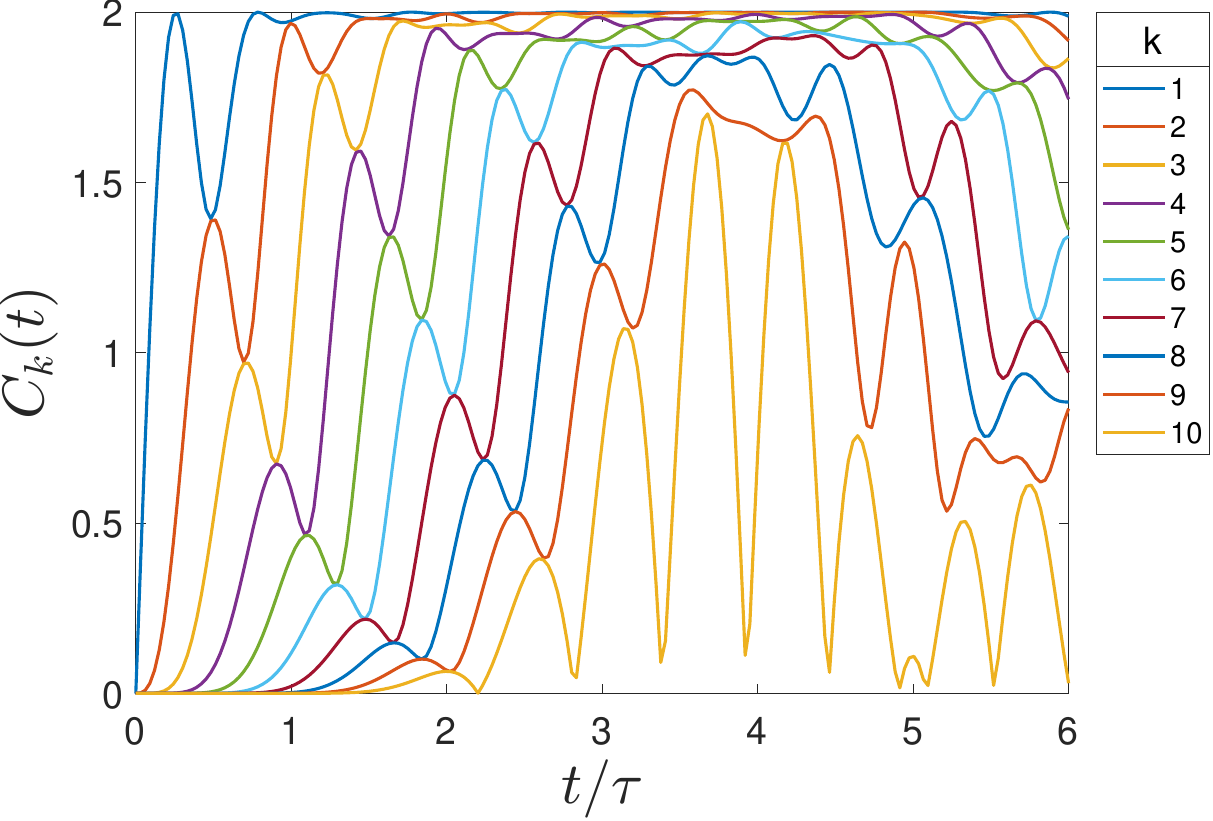}
    \caption{The Lieb--Robinson correlation function  for the TFIM of a ten-qubit chain with $J'=1/2$. This is the result of a direct calculation using the definition of \refeq{C_def_inh} and  \refeq{Z1Oft}.
    }
    \label{fig:DirectCforShortChains}
\end{figure}

For relatively short chains it is tractable to calculate $C_k(t)$ directly from \refeq{C_def_inh} and \refeq{Z1Oft}. This requires calculating the matrix exponential of the $2^{N_q}\times 2^{N_q}$ Hamiltonian matrix. The result for a ten-qubit line is shown in \reffig{DirectCforShortChains}. The correlation  initially turns on for each site as the quantum influence propagates down the chain. The effect of reflections from the end of the chain, however, quickly becomes apparent, and indeed dominant. The computational challenge here is the same as that for direct state-based calculations, namely that the size of the operators scales as $2^{N_q}$, so for even modestly large systems direct calculations like this become untenable. For example, Colmenarez and Luitz \cite{Luitz2020} use heroic methods to calculate systems with $N_q = 22$. 

Equations (\ref{eq:C_def_inh}) and \refeq{Z1Oft} provide the conceptual foundation for the
operator Pauli walk (OPW) method introduced in \cite{Lent2024} which addresses the problem of handling longer system sizes. A Pauli string is the tensor product of operators $I_k, X_k, Y_k$, or  $Z_k$ on each site; it is usual to suppress the identity operators notationally.
Each nested commutation of $\hat{H}'$ with $Z_1$ generates a new Pauli string, corresponding to a ``step'' in an abstract operator space whose nodes represent Pauli strings and whose edges encode their coupling under $\hat{H}'$. The space of Pauli strings has dimension $4^{N_q}$ and in general the whole space is required to support $Z_1(t)$ and therefore calculate $C_k(t)$. The OPW method exploits the fact that for the TFIM Hamiltonian $H'$, the nested commutations in \refeq{Z1Oft}  explore only a much smaller structured subspace of  size $2N_q$  forming a discrete walk that fully captures the dynamics.

To briefly illustrate, consider a four-qubit homogeneous chain. Starting from $Z_1$,  evaluating any number of  successive nested commutators with $H'_{4q}$  in \refeq{Z1Oft} generates only eight distinct Pauli strings:
\begin{eqnarray}
&\ & Z_1,\; Y_1,\; X_1 Z_2,\;  X_1 Y_2,\;  X_1 X_2 Z_3,\;\hfill\nonumber\\ 
&\ &   X_1 X_2 Y_3,\; X_1 X_2 X_3 Z_4,\; X_1 X_2 X_3 Y_4. \label{eq:4q_sigma_list}
\end{eqnarray}

\begin{figure}[!tbp]
    \centering
    \includegraphics[width=\figwidth]{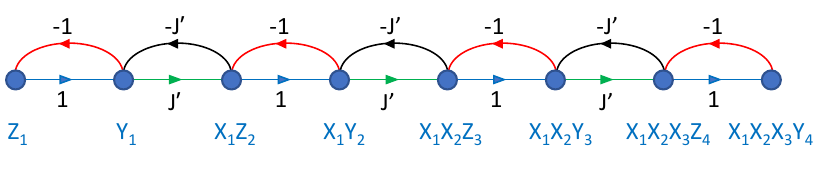}
    \caption{Pauli operator walk for a four-qubit TFIM chain.
        The nodes correspond to the eight operator strings
         listed in \eqref{eq:4q_sigma_list}, and the directed edges represent the value of commutators between $Z_1$
         and terms in $H'_{4q}$ that appear in \refeq{Z1Oft}
         }
    \label{fig:pauli_walk_graph}
\end{figure}

These eight operators form the nodes of  closed Pauli operator walks, shown schematically in Fig.~\ref{fig:pauli_walk_graph}.  The weight of each directed edge represents the value of a commutator in the expansion given by \refeq{Z1Oft}.  The sum in (\ref{eq:Z1Oft})  maps onto a sum over all walks  on the operator chain. 
Note that the walk is not along the qubit chain, but rather in this operator space. 
Thus, the nested commutators needed to calculate $Z_1(t)$ using \refeq{Z1Oft} under the Ising Hamiltonian explore only a compact, linearly--connected subset of the  full, exponentially larger, space of Pauli strings.

The operator walks can be described using an adjacency matrix which takes the  tridiagonal form
\begin{equation}
A_0 =
\begin{pmatrix}
0      & 1     & 0      & 0      & 0       & 0      & \cdots   & 0  \\
-1     & 0     & J'     & 0      & 0       & 0      & \cdots   & 0  \\
0      & -J'   & 0      & 1      & 0       & 0      & \cdots   & 0  \\
0      & 0     & -1     & 0      & J'      & 0      & \cdots   & 0  \\
0      & 0     & 0      & -J'    & 0       & \ddots & \vdots   & \vdots \\
0      & 0     & 0      & 0      & \ddots  & \ddots & J'       & 0  \\
\vdots & \vdots& \vdots & \vdots & \cdots  & -J'    & 0        & 1 \\
0      & 0     & 0      & 0      & \cdots  & 0      & -1       & 0  
\end{pmatrix}.
\label{eq:A_hom}
\end{equation}
As shown in \cite{Lent2024}, summing over all walks governed by this adjacency matrix yields the following closed-form expression for the LR correlation function on the homogeneous chain:
\begin{equation}
C^0_k(t) =
\sqrt{\sum_{m = 2k}^{2N_q}
\left|\, 2 \left[
\exp\!\left(-2 \pi \tfrac{t}{\tau}\,A_0\right)
\right]_{1,m} \right|^2 }.
\label{eq:C_eval_homogeneous}
\end{equation}

Evaluating \refeq{C_eval_homogeneous}  requires calculating  matrix exponentials of $\bm{A}_0$, a sparse real skew-symmetric matrix of size $2N_q \times 2N_q$. Using this approach, the LR correlation function can be calculated directly for any time without time-marching or Trotterization. Because \refeq{A_hom} scales only with $2N_q$, rather than $4^{N_q}$, this approach allows $C_k(t)$ to be calculated for chains of hundreds or even thousands of qubits. As we shall see below, for many problems a finite chain can be long enough that the results essentially produce those of a semi-infinite chain. 

\begin{figure}[!tbp]
    \centering
    \includegraphics[width=\figwidth]{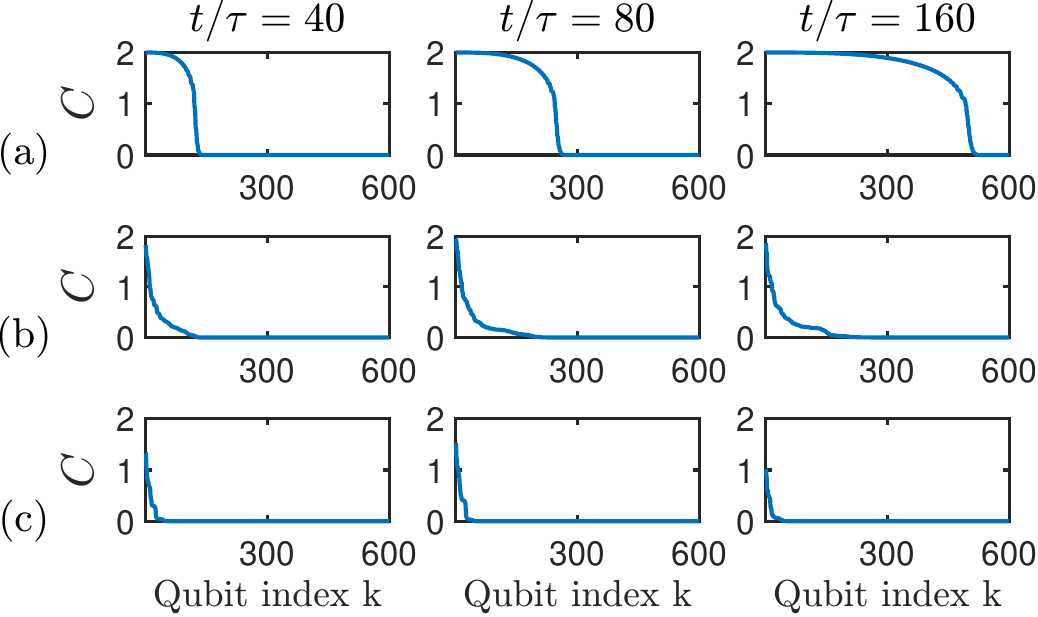}
    \caption{Snapshots of the time evolution of $C_k(t)$ for a  TFIM chain in the paramagnetic (weak-coupling) phase with $J'_0 = 0.5$. (a) No disorder, $\Delta J' = 0$. (b) Weak disorder, $\Delta J' = 0.4$. (c) Strong disorder, $\Delta J' = 0.8$. The chain is 1200 qubits long, but only the first 600 qubits are plotted.
    }
    \label{fig:LR_multiplot_WeakCoupling}
\end{figure}

Figure \ref{fig:LR_multiplot_WeakCoupling}(a) shows snapshots of the LR correlation function $C_k(t)$ calculated using the OPW method
at three different times. The Ising chain here is composed of $N_q=1200$ qubits, though the correlation function is shown only for the first 600. In this case, $J'=0.5$ so the system is in the paramagnetic phase. The front of quantum correlation moves smoothly down the line with the velocity given by \refeq{vFront} (i.e., $v=\pi$ qubits/$\tau$).

\section{Disordered Ising chains \label{sec:DisorderedIsing} }

We now generalize the homogeneous Ising chain to include spatially-varying couplings.
Working in the same dimensionless units, the inhomogeneous TFIM is
\begin{equation}
H' \;=\; - \sum_{k=1}^{N_q-1} J'_k \, Z_k Z_{k+1} \;-\; \sum_{k=1}^{N_q} X_k ,
\label{eq:H_dimless_inh}
\end{equation}
where $J'_k$ is the coupling between sites $k$ and $k+1$, which we again take to be non-negative but which can now vary down the chain.
As in the homogeneous case, the nested-commutator dynamics of $Z_1(t)$ given by \refeq{C_def_inh} is
confined to a $2N_q$-dimensional operator subspace rather than the full $2^{N_q}$-dimensional space of all Pauli strings. The only change is that
the operator-space adjacency matrix acquires position-dependent weights:
\begin{equation}
A \;=\;
\begin{pmatrix}
0      & 1     & 0      & 0      & 0       & 0      & \cdots   & 0  \\
-1     & 0     & J_1'   & 0      & 0       & 0      & \cdots   & 0  \\
0      & -J_1' & 0      & 1      & 0       & 0      & \cdots   & 0  \\
0      & 0     & -1     & 0      & J_2'    & 0      & \cdots   & 0  \\
0      & 0     & 0      & -J_2'  & 0       & \ddots & \vdots   & \vdots \\
0      & 0     & 0      & 0      & \ddots  & \ddots & J_{N_q-1}' & 0  \\
\vdots & \vdots& \vdots & \vdots & \cdots  & -J_{N_q-1}' & 0    & 1 \\
0      & 0     & 0      & 0      & \cdots  & 0      & -1       & 0  
\end{pmatrix}.
\label{eq:A_inh}
\end{equation}
Thus, the solution to the inhomogeneous problem requires only a straightforward substitution
$\bm{A}_0 \!\to\! \bm{A}$. We sum over walks on the same set of operators; only the edge weights of the Pauli walk change.
The LR correlation function therefore retains the same closed form, just in terms of the matrix $\bm{A}$:
\begin{equation}
C_k(t) \;=\;
2\sqrt{\sum_{m = 2k}^{2N_q}
\left|\, \left[
\exp\!\left(-2 \pi \tfrac{t}{\tau}\,A\right)
\right]_{1,m} \right|^2 }.
\label{eq:C_eval_inh}
\end{equation}

Within this inhomogeneous framework, we construct specific instances of disordered chains by
independently sampling each coupling $J'_k$ from a 
probability distribution.
We use a uniform probability density centered at a baseline coupling $J'_0$
with width $\Delta J'$:
\begin{equation}
P(J')=
\begin{cases}
\dfrac{1}{\Delta J'} & \text{for } J' \in 
\left[J'_0 - \tfrac{\Delta J'}{2},\, J'_0 + \tfrac{\Delta J'}{2}\right],\\[4pt]
0 & \text{otherwise}.
\end{cases}
\label{eq:UniformPDF_inh}
\end{equation}

\reffig{multi_panel_ising}(c) illustrates the
 disorder distributions used in this paper. For the paramagnetic phase, we consider two cases centered at $J'_0 = 0.5$: a weak-disorder distribution with
$\Delta J' = 0.4$ (blue) and a strong-disorder distribution with
$\Delta J' = 0.8$ (red). For the ferromagnetic phase, we use two analogous
distributions centered at $J'_0 = 1.5$, with $\Delta J' = 0.4$ (black) and
$\Delta J' = 0.8$ (green).

For the case we consider here, with disorder in the couplings only, we define a simplified Ising parameter \cite{Fisher1995,Igloi2005}
\begin{equation}
    \delta\equiv\left[ \log(J')\right]_{avg}.
    \label{eq:deltaDefine}
\end{equation}
In the presence of disorder, the critical point is not at $J'=1$, but rather at $\delta=0$, usually corresponding to a value of $J'$ slightly above 1.  The  parameter $\delta$ is negative for the paramagnetic phase and  is positive for the ferromagnetic phase. 


For the uniform distribution function in \refeq{UniformPDF_inh}, we can calculate $\delta$ explicitly in terms of $J'_0$ and $\Delta J'$, yielding
\begin{align}
  \delta&= \frac{J'_0}{\Delta J'} \log\left( \frac{J'_0 + {\Delta J'}/{2}}{J'_0 - {\Delta J'}/{2} }\right) \nonumber\\
  &+ \frac{1}{2} \log\left( (J'_0)^2 - \frac{(\Delta J')^2}{4} \right) - 1  .
\end{align}
The parameter $\delta$ then varies strongly with  $J'_0$ for a fixed value of $\Delta J'$, but only weakly with $\Delta J'$ for fixed $J'_0$.

In the following sections, we calculate the disorder-averaged LR correlation function  by considering $N_c$ independently-sampled configurations of the random couplings $\{J'_k\}^{(m)}$, where $m$ is the index of a particular configuration of couplings. Then the average value of the correlation function $\overline{C}$ is given by
\begin{equation}
\overline{C_{k}}(t) \;\equiv\; \frac{1}{N_c} 
\sum_{m=1}^{N_c} C_{k}^{(m)}(t)
\label{eq:AveragedC_inh}
\end{equation}
where each $C_{k}^{(m)}(t)$ is evaluated from
\refeq{C_eval_inh} for each particular sample.  We  similarly define the \emph{typical} value $C^{typ}$ as the geometric mean  over the $N_c$ sample configurations.  

\begin{equation}
    C^{typ}(k,t)\equiv \exp\left[ \frac{1}{N_c} \sum_{m=1}^{N_c} \log\left(C^{(m)}_k(t)\right) \right]
    \label{eq:PRtyp}
\end{equation}
\noindent Importantly, these two types of averages may yield quite different results in some circumstances \cite{Igloi2005}.


\subsection{Paramagnetic phase \label{ParamagneticPhase}}
We  focus first on the results for the LR correlation function
$C_k(t)$ in the paramagnetic (weak-coupling) phase, taking a baseline coupling of
$J'_0 = 0.5$. 

Figure \ref{fig:LR_multiplot_WeakCoupling}(b) shows snapshots of the LR correlation function $C_k(t)$ on a 1200-qubit chain (only the first 600 qubits are shown) for a specific random distribution of couplings drawn from \refeq{UniformPDF_inh} with $J'_0=0.5$  and $\Delta J'= 0.4$ ($\delta= -0.7212$). Figure \ref{fig:LR_multiplot_WeakCoupling}(c) shows the case of $\Delta J'= 0.8$ ($\delta= -0.8307$). Increasing disorder in the coupling between qubits clearly impedes propagation of  quantum correlations down the chain.

\begin{figure}[!tbp]
  \centering
  \begin{minipage}{\figwidth}
    \makebox[16pt][l]{\textbf{(a)}\hspace{1em}}%
    \includegraphics[width=\linewidth]{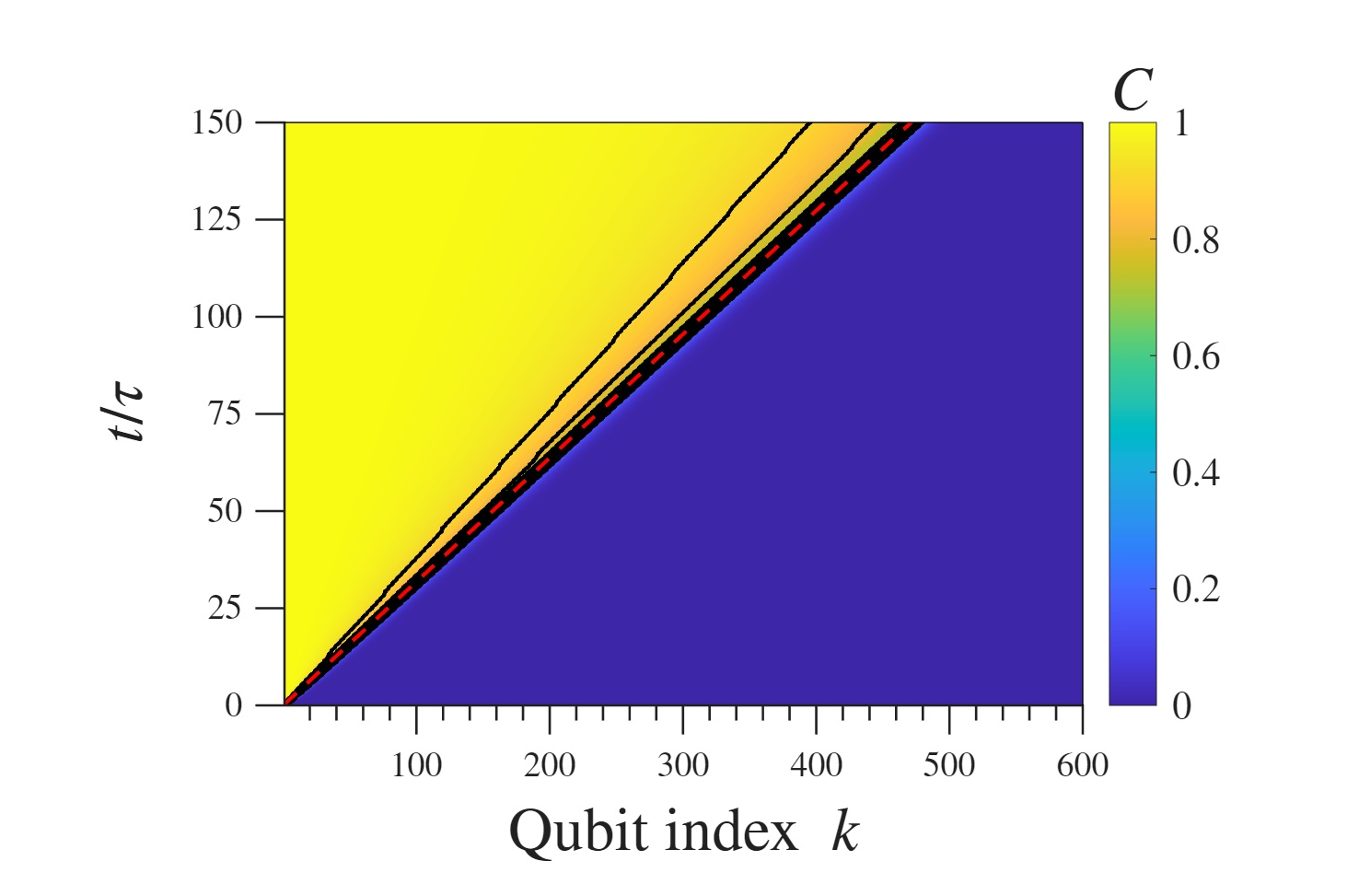}
  \end{minipage}

  \vspace{1em}

  \begin{minipage}{\figwidth}
    \makebox[16pt][l]{\textbf{(b)}\hspace{1em}}%
    \includegraphics[width=\linewidth]{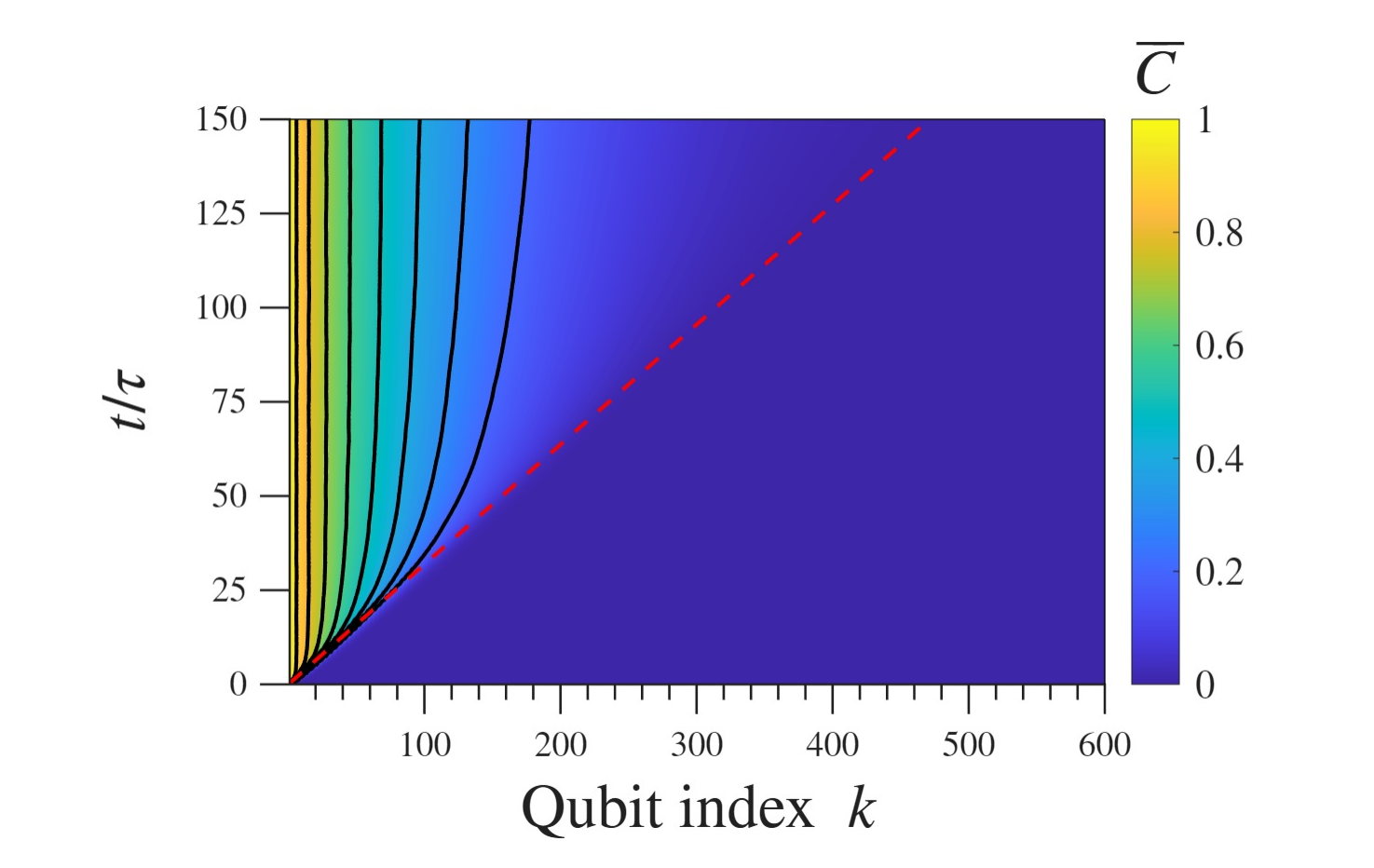}
  \end{minipage}

  \begin{minipage}{\figwidth}
    \makebox[16pt][l]{\textbf{(c)}\hspace{1em}}%
    \includegraphics[width=\linewidth]{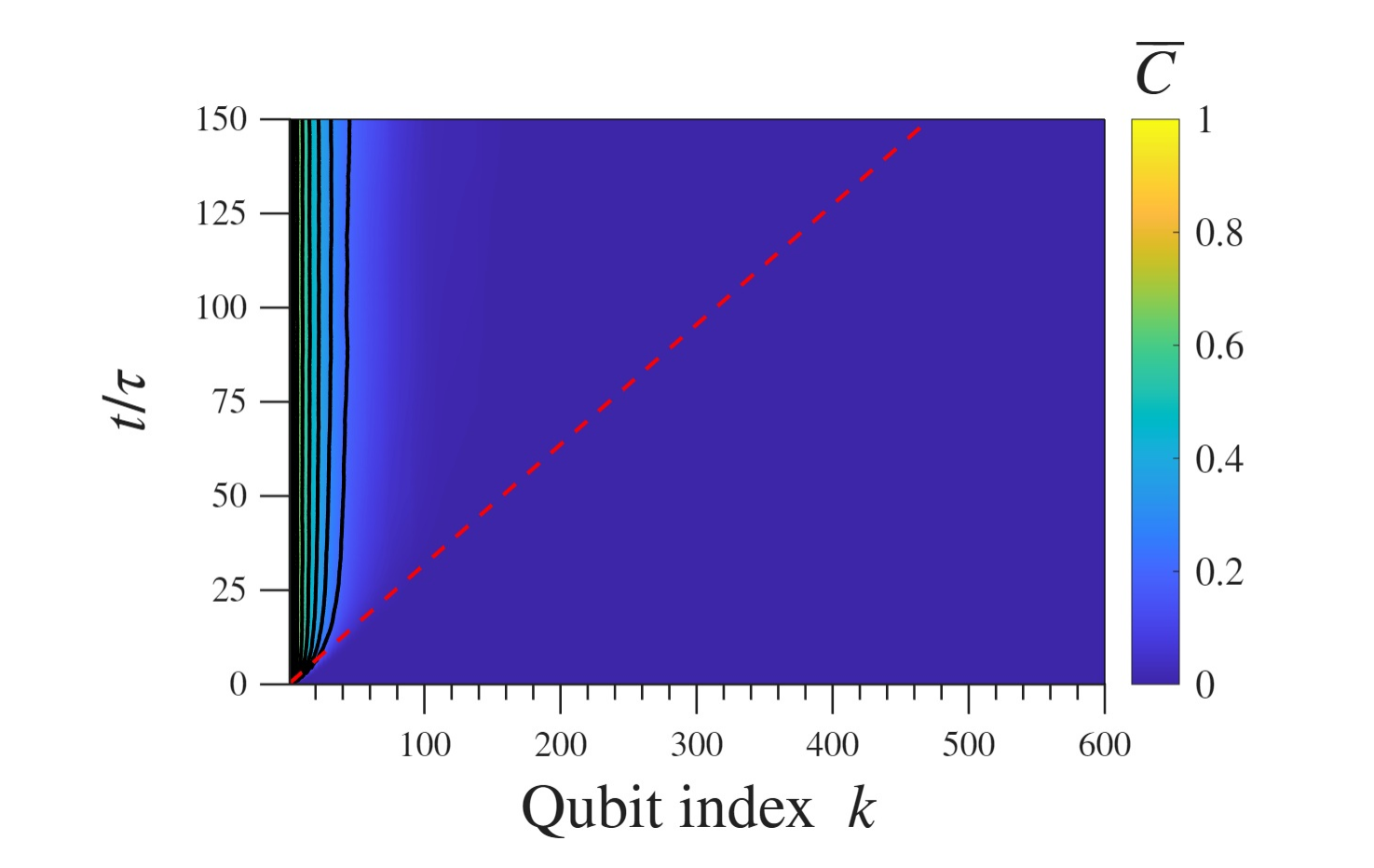}
  \end{minipage}

  \caption{Light-cone diagrams for Ising chain in the paramagnetic phase, $J' = 0.5$, with (a) no disorder, $\Delta J' = 0$, 
  (b) weak disorder, $\Delta J' = 0.4$, and 
  (c) strong disorder, $\Delta J' = 0.8$. In (b) and (c) the disorder-averaged mean value of the Lieb-Robinson correlation function $\overline{C}$ is plotted.
  }
  \label{fig:ParaLightConePlots}
\end{figure}

\reffig{ParaLightConePlots}(a) shows the light-cone plot for the reference situation with no disorder, $\Delta J'=0$, and $J'_0=0.5$. The color of the plot represents the value of $C_k(t)$ for this chain of 1200 qubits (600 are shown). Isocontours of $C$ are shown for $C=[1, 0.9, 0.8, 0.7, 0.6, 0.5, 0.4]$. They are closely spaced together in this plot, but are indicated for comparison with the disordered cases below. The red dashed line corresponds to the propagation velocity given by \refeq{vFront}. (Note that the slope of the line is the inverse of the velocity.) The light-cone plot is a fuller visualization of the snapshots of the moving correlation front in  \reffig{LR_multiplot_WeakCoupling}(a). 

\FloatBarrier

\reffig{ParaLightConePlots}(b) shows the light-cone plot for a 600 qubit disordered Ising chain with $J'_0=0.5$ and $\Delta J'=0.4$ ($\delta= -0.7212$). Here the color represents  the configuration-averaged LR correlation function $\overline{C}$ defined in \refeq{AveragedC_inh}, averaged over $N_c=500$ configurations. The probability density from which the values of $J'_k$ are drawn is shown schematically in \reffig{multi_panel_ising}(c)  in red. Isocontours are shown for the values of $\overline{C}=[1, 0.9, 0.8, 0.7, 0.6, 0.5, 0.4]$. The dashed red line in the light-cone plot again represents the velocity of the front in the disorder-free case.  Close to the origin, the $\overline{C}$ isocontours  follow the disorder-free velocity line, but they soon bend upward and the front velocity slows. As time progresses, the isocontours become closer and closer to vertical.   This signifies that the front has stopped advancing: correlations become stationary in space, marking the onset of localization.

The high-disorder light-cone, with $\Delta J'=0.8$ ($\delta= -0.8307$), is shown in \reffig{ParaLightConePlots}(c). The probability density from which the values of $J'_k$ are drawn is shown as the red curve in \reffig{multi_panel_ising}(c). The $\overline{C}$ isocontours in this case   almost immediately turn vertical, corresponding to no propagation of the correlation front.

The light-cone diagrams in \reffig{ParaLightConePlots}(b) and (c) display the mean correlation function $\overline{C}$. At the scale of the diagrams, similar plots of the geometric mean $C^{typ}$ would be indistinguishable from those shown. That is true for the ferromagnetic regime discussed in the next section as well.  We will examine the difference between the two means in section \ref{sec:CriticalRegion} below.

\begin{figure}[!tbp]
    \centering
    \includegraphics[width=\figwidth]{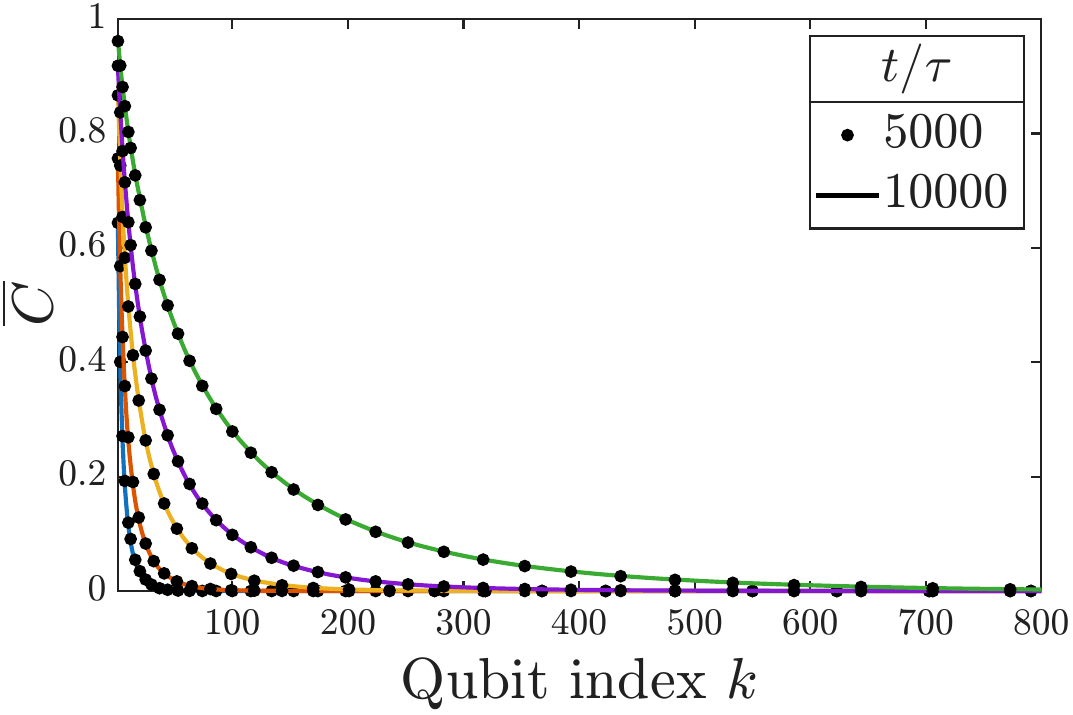}
    \caption{Disorder-averaged LR correlation function in the paramagnetic phase for  two late times. The TFIM chain is 1000 qubits in length with  baseline coupling $J'_0 = 0.5$.  $\overline{C}(k,t)$ is averaged over 1000 disorder configurations.         Results are shown at $t/\tau =$ 5,000 and $t/\tau =$ 10,000 for 
        different disorder strengths: 
        $\Delta J' = 0.2$ (green),  $\Delta J' = 0.3$ (purple), 
        $\Delta J' = 0.4$ (yellow), $\Delta J' = 0.6$ (red), 
        and $\Delta J' = 0.8$ (blue).  
        The identity of the two late-time curves (dots and lines) 
        indicates that the penetration of quantum correlations down the chain in each case has become stationary.}
    \label{fig:paramagnetic_decay_plot}
\end{figure}

The light-cone plots in \reffig{ParaLightConePlots} extend to time $t/\tau=150$. If many-body localization has  stopped the propagation of quantum correlations down the chain, we would expect to see the correlation function at longer times become stationary, \emph{i.e.}, independent of time. We can check this by examining $\overline{C}$ at two much longer times. \reffig{paramagnetic_decay_plot} shows the disorder-averaged LR correlation function for times $t/\tau$=5,000 (dots) and $t/\tau=$10,000 (solid lines). Here we use a chain of $N_q$=1000 qubits and  average over $N_c=$1000 disorder configurations. The results are shown for various amounts of disorder: $\Delta J' =[0.2, 0.4, 0.6, 0.8]$ corresponding to  $\delta \in [-0.83,-0.69]$. The fact that the dots fall on the lines indicates that in each case the dynamics of the system has localized quantum correlations that are no longer moving down the chain---zero propagation velocity. 

\begin{figure}[!tbp]
    \centering
    \includegraphics[width=\figwidth]{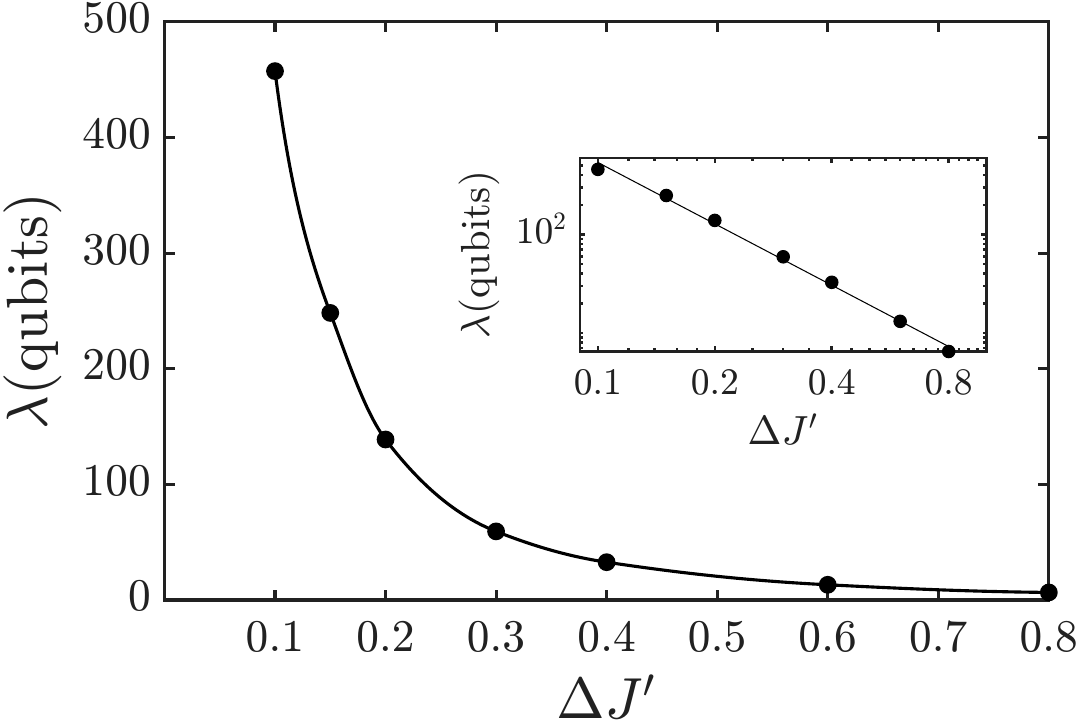}
    \caption{
    Correlation length as a function of the coupling disorder.
        The system is 1000 qubits long with $J'_0=0.5$ and the disorder is averaged over $N_c=$1000 configurations. The analysis is for a late time, here $t/\tau=$10,000, by which point the LR correlation function has become stationary, as shown in \reffig{paramagnetic_decay_plot}.
        For each disorder strength $\Delta J'$, 
        $\lambda$ is the 
         distance (in qubits) for which
          $\overline{C}_k$ drops below a threshold value, here 0.2.
        The decrease of $\lambda$ with $\Delta J'$  
        quantifies the
        increasing spatial confinement of correlations as disorder grows.
        }
    \label{fig:paramagnetic_decay_length}
\end{figure}

In both the light-cone plots of \reffig{ParaLightConePlots} and the later-time curves of \reffig{paramagnetic_decay_plot}, it is clear that higher amounts of disorder (larger $\Delta J'$) result in a more rapid decay of correlations down the chain. For example, in the $\Delta J' =0.1$ case  shown in \reffig{paramagnetic_decay_plot}, $\overline{C}$ is substantial out to several hundred qubits, whereas when  $\Delta J' =0.8$, it is extinguished a few tens of qubits down the chain. This observation can be made quantitative by calculating the qubit index for which $\overline{C}$ first drops below a threshold level for a late time, here chosen to be $t_{\text{late}}/\tau=$10,000. We define $C_{\text{thresh}}$ as the threshold level (we use 0.2 below) and the corresponding qubit index $\lambda$ ,
\begin{equation}
\lambda (\Delta J') \;\equiv\; 
\min\Big\{\,k :\; \overline{C}_k(t_{\mathrm{late}}) \le C_{\text{thresh}} \Big\}.
\label{eq:k_threshold}
\end{equation}

\noindent   The quantity $\lambda$  then quantifies an effective  localization length for the correlations. \reffig{paramagnetic_decay_length}  shows $\lambda$  as a function of
the disorder strength $\Delta J'$. The values are derived from the same 1000-qubit chain described in \reffig{paramagnetic_decay_plot}. 

The inset in \reffig{paramagnetic_decay_length} shows a log-log plot of $\lambda$ vs. $\Delta J'$. The fitted slope of $-2.07$ is consistent with a $\lambda \sim 1/(\Delta J')^2$ dependence.

\subsection{Ferromagnetic phase \label{FerromagneticPhase}}

\begin{figure}[!tbp]
    \centering
    \includegraphics[width=\figwidth]{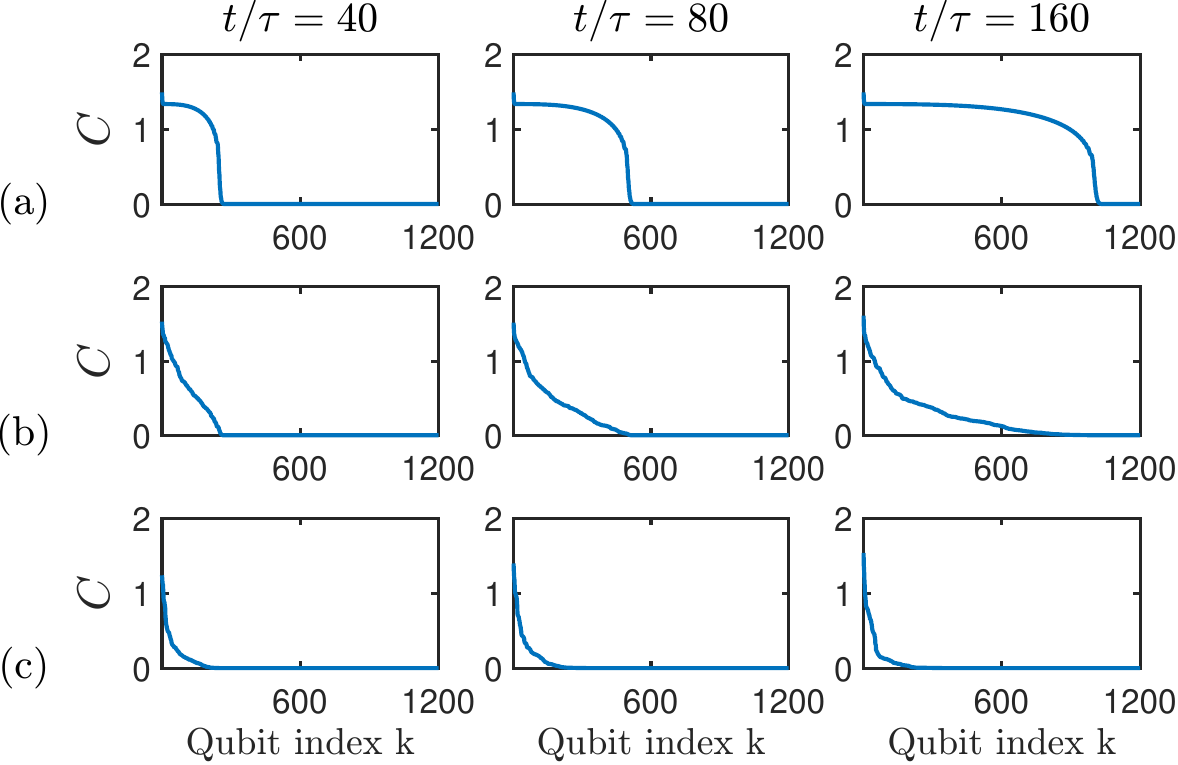}
    \caption{
    Snapshots of the time evolution of $C_k(t)$ for a  TFIM chain in the ferromagnetic (strong-coupling) phase with $J'_0 = 1.5$. (a) No disorder, $\Delta J' = 0$. (b) Weak disorder, $\Delta J' = 0.4$. (c) Strong disorder, $\Delta J' = 0.8$. 
 }
    \label{fig:LR_multiplot_ferromagnetic}
\end{figure}

\reffig{LR_multiplot_ferromagnetic}(a) shows snapshots of the LR correlation function at three times for a 1200-qubit TFIM chain  in the ferromagnetic phase, $J_0'=1.5$, with no disorder, $\Delta J'=0$. 
The  quantum correlation front moves smoothly down the line with the velocity given by \refeq{vFront} (i.e., $v=2\pi$ qubits/$\tau$), faster than in the paramagnetic case shown in \reffig{LR_multiplot_WeakCoupling}(a). 

\reffig{LR_multiplot_ferromagnetic}(b) and (c) show snapshots of $C_k(t)$ for the same length chain and $J'_0=1.5$, but with coupling disorder characterized  by  $\Delta J'=0.4$ ($\delta= 0.4025$) \ and $\Delta J'=0.8$ ($\delta= 0.3934$), respectively. These are each shown for one specific disorder configuration, drawn independently from the probability distribution of \refeq{UniformPDF_inh}. We see that in the ferromagnetic phase, not only is the propagation faster than in the paramagnetic phase, but the quantum correlation penetrates down the chain farther. Higher disorder still impedes propagation and localizes the correlated region nearer the origin.

\begin{figure}[!tbp]
  \centering
  \begin{minipage}{\figwidth}
    \makebox[16pt][l]{\textbf{(a)}\hspace{1em}}%
    \includegraphics[width=\linewidth]{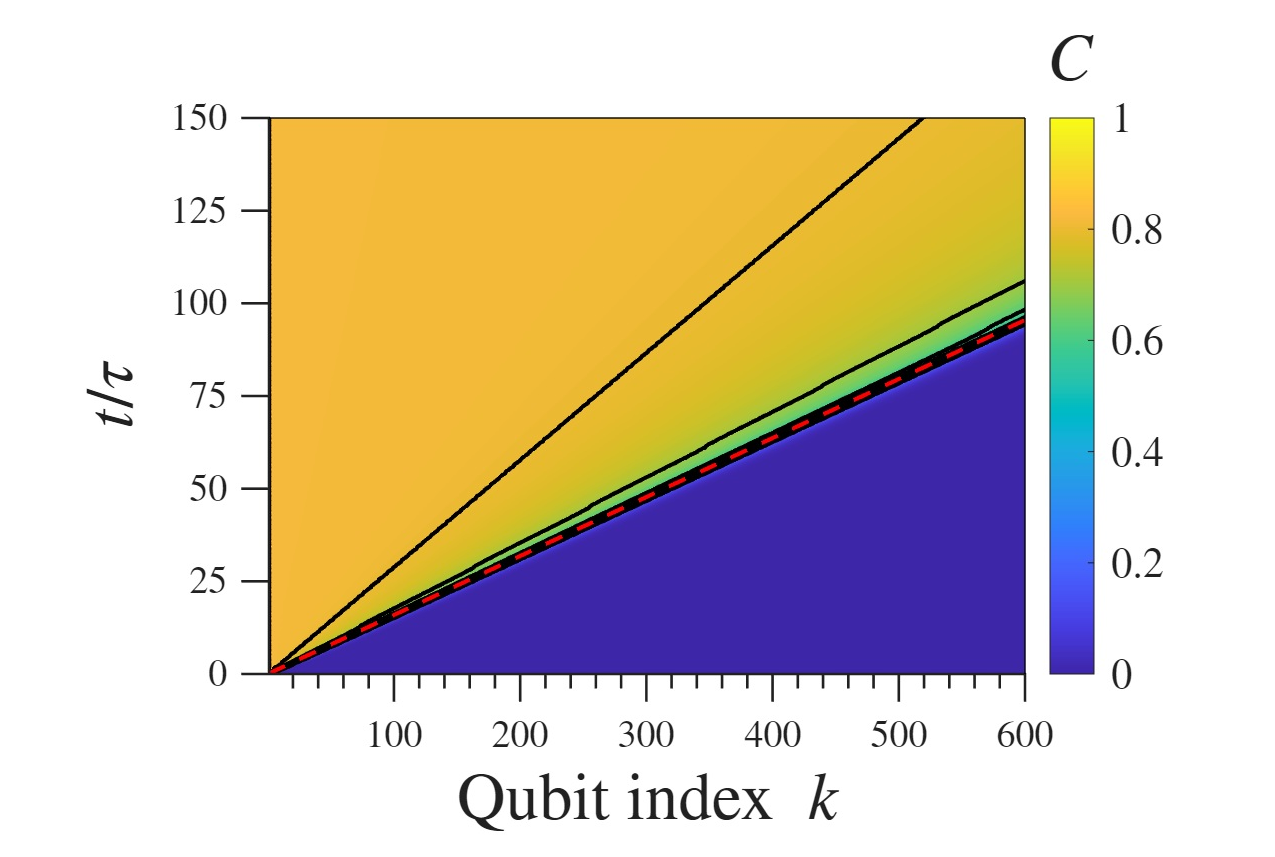}
  \end{minipage}

  \vspace{1em}

  \begin{minipage}{\figwidth}
    \makebox[16pt][l]{\textbf{(b)}\hspace{1em}}%
    \includegraphics[width=\linewidth]{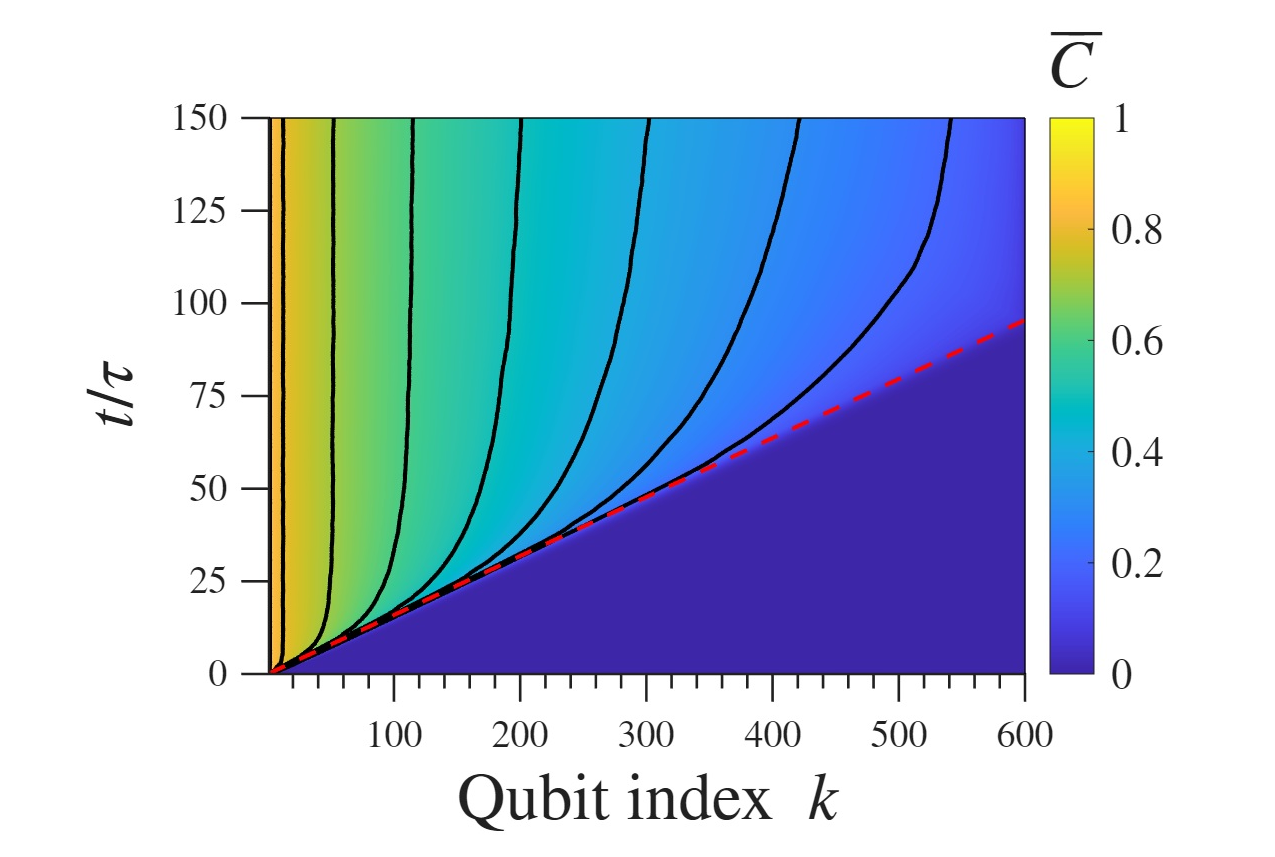}
  \end{minipage}

  \begin{minipage}{\figwidth}
    \makebox[16pt][l]{\textbf{(c)}\hspace{1em}}%
    \includegraphics[width=\linewidth]{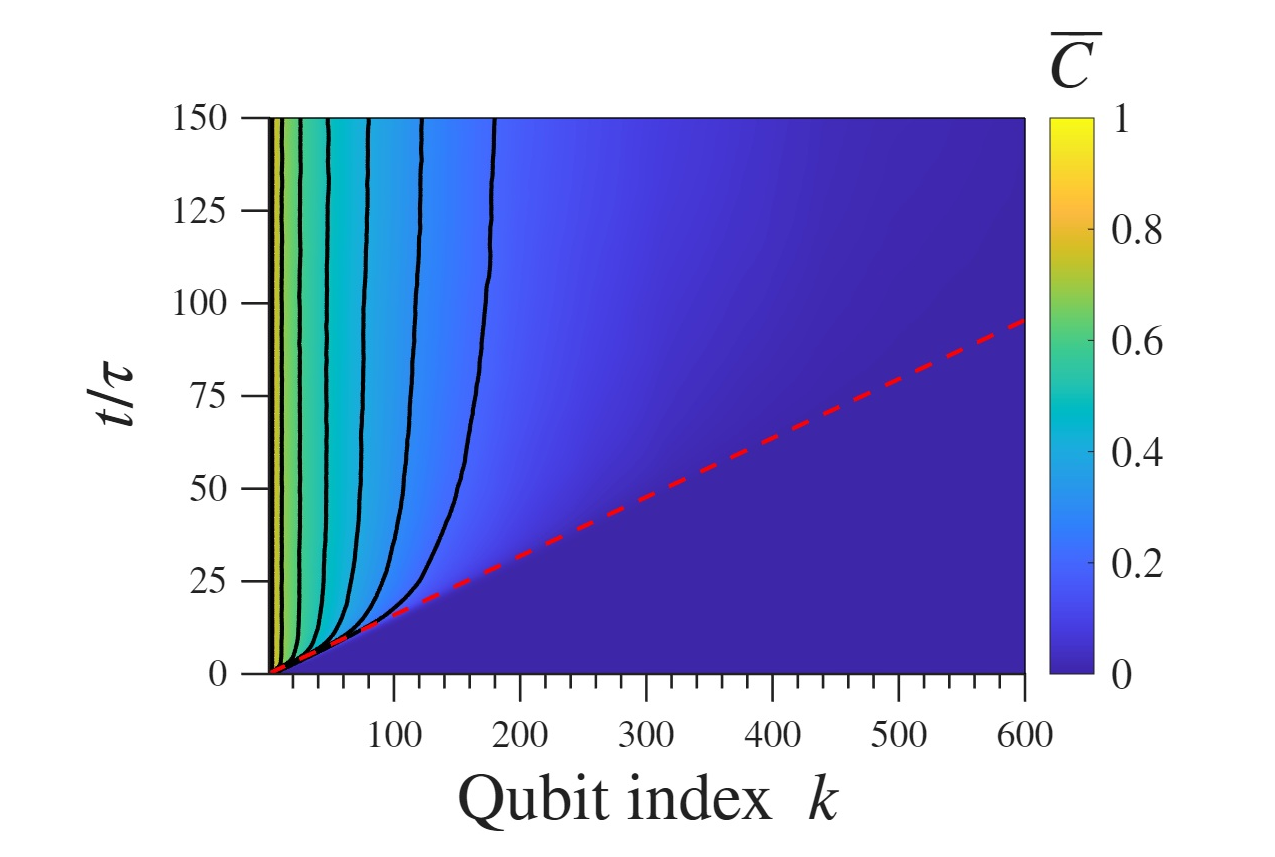}
  \end{minipage}

  \caption{Light-cone plots for Ising chain in the ferromagnetic phase, $J'=1.5$  with (a) no disorder, $\Delta J' = 0$, 
  (b) weak disorder, $\Delta J' = 0.4$, and 
  (c) strong disorder, $\Delta J' = 0.8$. 
  In (b) and (c) the disorder-averaged mean value of the Lieb-Robinson correlation function $\overline{C}$ is plotted.
  }
  \label{fig:FerroLightConePlots}
\end{figure}

\reffig{FerroLightConePlots}(a) shows the light-cone plot of $C_k(t)$ for a disorder-free 600-qubit chain in the ferromagnetic phase with $J_0'=1.5$. The color map represents the magnitude of the LR correlation function, and the overlaid isocontours are for levels $C=[1, 0.9, 0.8, 0.7, 0.6, 0.5, 0.4]$, though they are hard to separate visually. The red dashed line corresponds to the correlation front velocity of \refeq{vFront}.

\FloatBarrier

\reffig{FerroLightConePlots}(b) and (c) are light-cone plots of the disorder-averaged LR correlation function $\overline{C}_k(t)$ with $N_c=500$. As above, $J'_0=1.5$ and $\Delta J'=0.4$ for the lower-disorder case, with $\Delta J'=0.8$ for the higher-disorder case, corresponding to $\delta=0.4025$ and $0.3934$. Isocontours of $\overline{C}$ for the same levels as above are indicated.
The red dashed line shows the front velocity for the disorder-free case. These light-cones are qualitatively similar to those for the paramagnetic case shown in \reffig{ParaLightConePlots}. In the absence of disorder, the correlation front moves ballistically down the chain, yielding straight light-cones. Disorder in the couplings causes the light-cones to bend upward and become vertical.

The group velocity is larger in the ferromagnetic phase than in the paramagnetic phase, as shown in \reffig{multi_panel_ising}(b).
As a result correlations are able to penetrate greater distances down the chain  before being attenuated by scattering due to the disorder. The velocity does not further increase with $J'_0$, so the penetration down the chain does not continue to increase deeper into the ferromagnetic regime ($J'_0>>1)$.



\begin{figure}[!tbp]
    \centering
    \includegraphics[width=\figwidth]{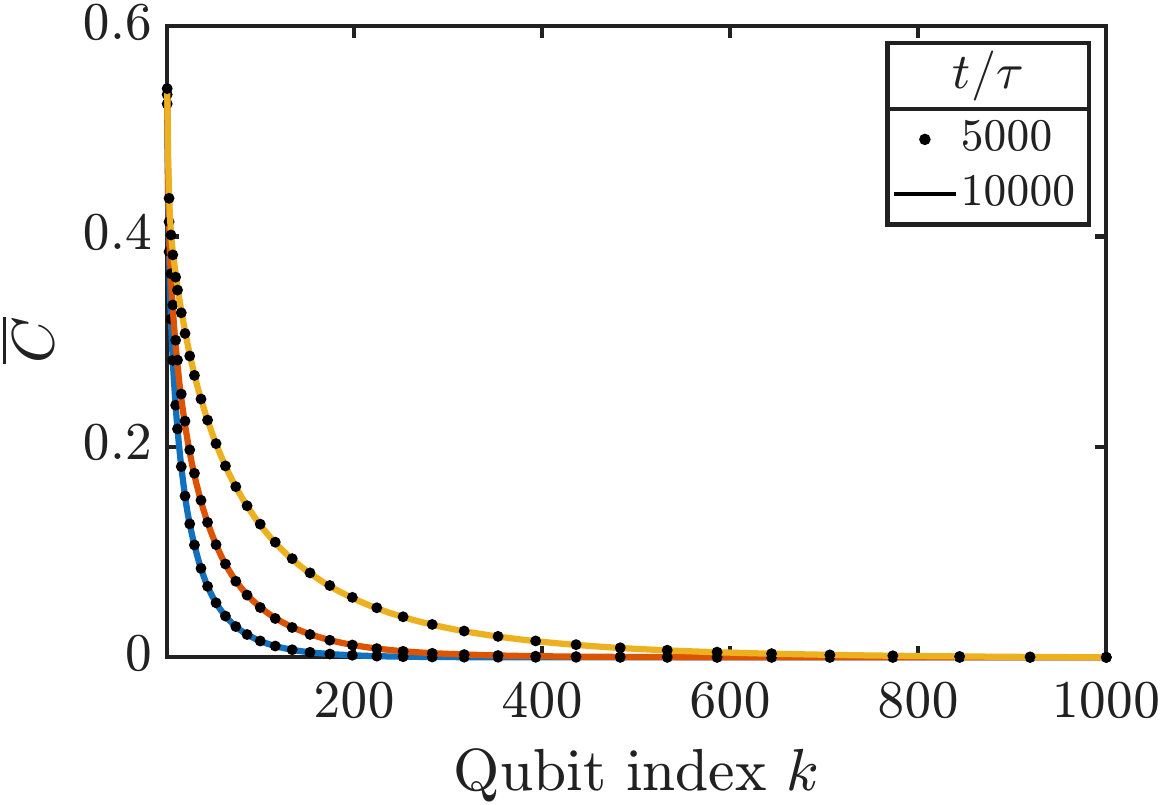}
    \caption{
Disorder-averaged LR correlation function in the ferromagnetic phase for  two late times. The TFIM chain is 1000 qubits in length with  baseline coupling $J'_0 = 1.5$.  $\overline{C}(k,t)$ is averaged over 1000 disorder configurations.         Results are shown at $t/\tau =$ 5,000 and $t/\tau =$ 10,000 for 
        different disorder strengths: 
        $\Delta J' = 0.4$ (yellow), $\Delta J' = 0.6$ (red), 
        and $\Delta J' = 0.8$ (blue).  
        The identity of the two late-time curves (dots and lines) 
        indicates that the penetration of quantum correlations down the chain in each case has become stationary. Compared to the weak-coupling (paramagnetic) case shown in \reffig{paramagnetic_decay_plot}, correlations penetrate farther down the chain, corresponding to a larger localization length in the ferromagnetic phase.
}
    \label{fig:ferromagnetic_decay_plot}
\end{figure}

As in the paramagnetic case, we can compare the LR correlation function at two late times to confirm that the propagation of quantum correlations has indeed ceased. \reffig{ferromagnetic_decay_plot} shows the disorder-averaged
correlation function $\overline{C}(k,t)$ at
$t/\tau=$  5,000 (dots) and $t/\tau =$ 10,000 (solid line) for a 1000-qubit system,
averaged over $N_c = $1000 disordered configurations, with
$\Delta J' = [0.4, 0.6, 0.8]$, ($\delta \in [0.4025,0.3934]$).
The results indicate that the
spatial profile has reached its stationary form.
Again, it is clear that relative to the paramagnetic results
of \reffig{paramagnetic_decay_plot},
the ferromagnetic correlations decay more slowly in $k$ at each $\Delta J'$,
indicating that correlations extend farther from the origin and the effective
localization length is larger in the strong-coupling phase.

Hamza, Sims, and Stolz (HSS) found that under certain conditions, the LR correlation function for the disordered XY model (of which the Ising model is a special case) has a stationary exponential bound 
\begin{equation}
    C_k(t) \le K e^{-\alpha (k-1)},
    \label{eq:HSS_bound}
\end{equation}
where $\alpha$ is a characteristic
decay constant \cite{Hamza2012, Stolz2017}.  
To compare this bound to our result, we evaluate the stationary disorder-averaged correlation profile $\overline{C}_k(t_{\mathrm{late}})$ at  time $t_{\mathrm{late}}/\tau=$10,000.  
Figures \ref{fig:paramagnetic_decay_plot} and \ref{fig:ferromagnetic_decay_plot}
established that by this time the  correlation front has ceased to propagate, so we can use $\overline{C}_k$ at $t_{\text{late}}$ to represent the stationary correlation profile $\overline{C}(\infty)$.  

We set $K=\overline{C}_1(t_{\mathrm{late}})$ and then determine the largest decay constant $\alpha$ for
which the inequality in \refeq{HSS_bound} holds for all qubit indices
$k$. That is, we find the $\alpha$  which  optimally bounds our calculated result.

\begin{figure}[!tbp]
    \centering
    \includegraphics[width=\figwidth]{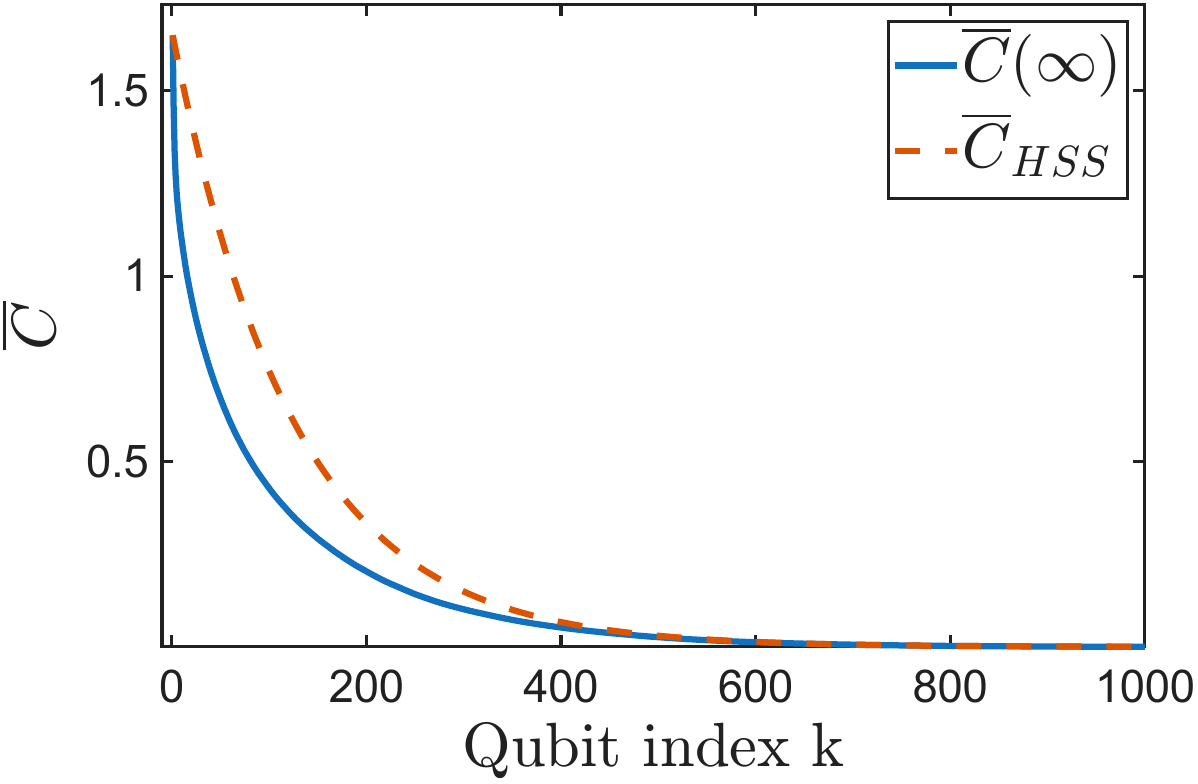}
    \caption{
        Comparison between the disorder-averaged late-time LR correlation function
        and the exponential bound of Hamza, Sims, and Stolz (HSS) 
        [\refeq{HSS_bound}]. The parameters for the calculation are $J'_0=1.5$, $\Delta J'=0.8$, ($\delta=0.3934)$, and $N_c=$1000  
        }
    \label{fig:ferromagneticBoundcheck}
\end{figure}

\reffig{ferromagneticBoundcheck} compares this exponential bound
with our  results for a 1000-qubit chain in the ferromagnetic phase with $J'_0=1.5$, $\Delta J'=0.8$,\ ($\delta = 0.3934$)\  and $N_c=$1000.
These results conform to the HSS bound, but it is noticeable that the calculated correlations decay significantly faster. This holds for both strong and weak
disorder and for both paramagnetic and ferromagnetic phases.

\subsection{Critical point region \label{sec:CriticalRegion}}

In many respects, the behavior around the critical point $\delta=0$ is very similar to that shown in Sections A and B above. The propagation behavior of Lieb-Robinson correlations largely varies smoothly as $\delta$ is varied across the quantum phase transition,  so we do not produce them here. Light-cone diagrams shift smoothly from the behavior seen in the paramagnetic phase to that of the ferromagnetic phase. This is not unexpected since the correlation front moves at the speed given  by \refeq{vFront} and illustrated in \reffig{multi_panel_ising}, a smoothly varying function. 

Around the critical point it is well known that Griffiths phases occur in the random TFIM \cite{Griffiths1969, RiegerYoung1996, IgloiKovacs2022}.  These are rare  regions with significant lengths of ferromagnetic order embedded in paramagnetic bulk regions, or vice versa.  The occurrence of these phases can result in singularities in the scaling behavior of various equilibrium  (or near-equilibrium) quantities such as the energy gap and the average ground state expectation value of the surface magnetization \cite{Igloi2005}.  One result is  that the behavior of average quantities is sometimes quite distinct from typical  quantities. 

\begin{figure}[!tbp]
    \centering
    \includegraphics[width=\figwidth]{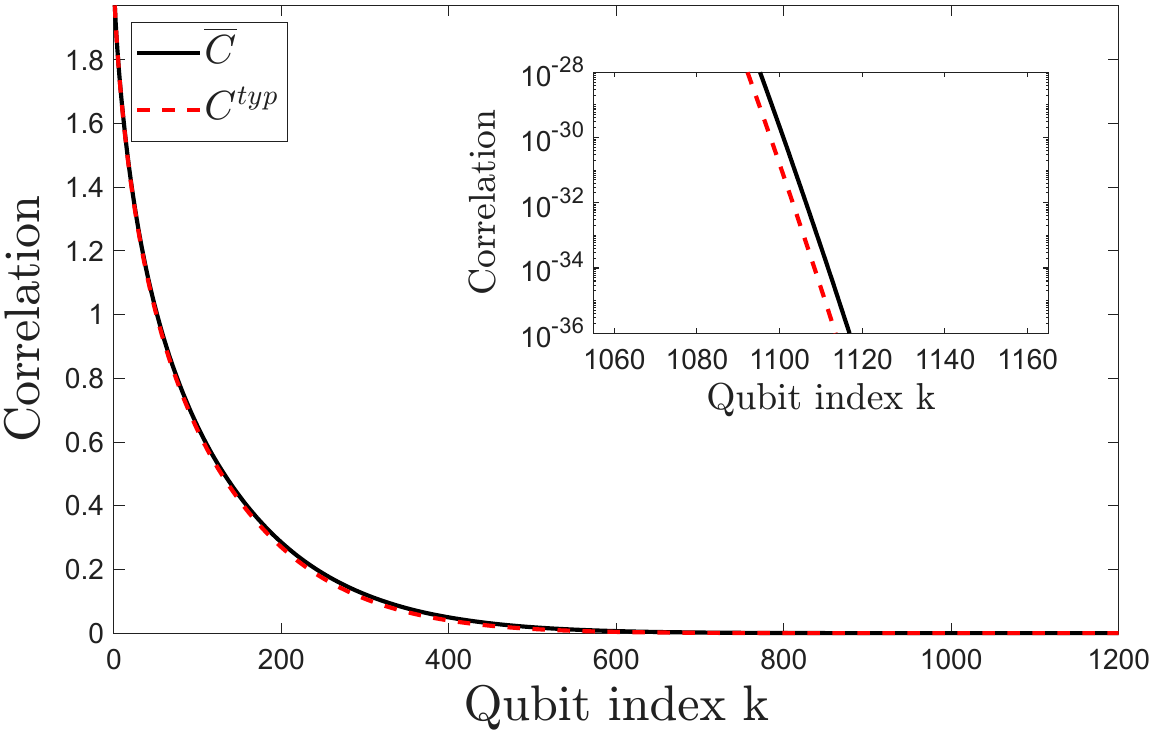}
    \caption{
        The sample-averaged Lieb-Robinson correlation function for disorder that is centered on the critical point. 
        Here $N_q=1200$, $J'_{0}=1.0151$ , $\Delta J'= 0.6$ ($\delta=0$), $t=160\tau$, and $N_{c}=20,000$. The arithmetic average $\overline{C}$ and the geometric average $C^{typ}$ are nearly identical when the correlation is large, but are significantly different for very small values deeper in the chain.  
        }
    \label{fig:CorrelationsNearCritical}
\end{figure}

Can we see evidence of this divergence between the two means? Figure \ref{fig:CorrelationsNearCritical} compares  the average Lieb-Robinson correlation function $\overline{C}_k$ with $C^{typ}_k$ in the region around the critical point $\delta=0$ ($J'\approx 1.015$) with $\Delta J'=0.6$. 
The averages are over a set of $N_c=20,000$ samples and the system has $N_q=1200$ qubits.  The time  $t=160\tau$ is chosen so that the system is essentially in its steady-state, i.e.,  propagation of correlations down the chain has stopped. 

\begin{figure}[!tbp]
    \centering
    \includegraphics[width=\figwidth]{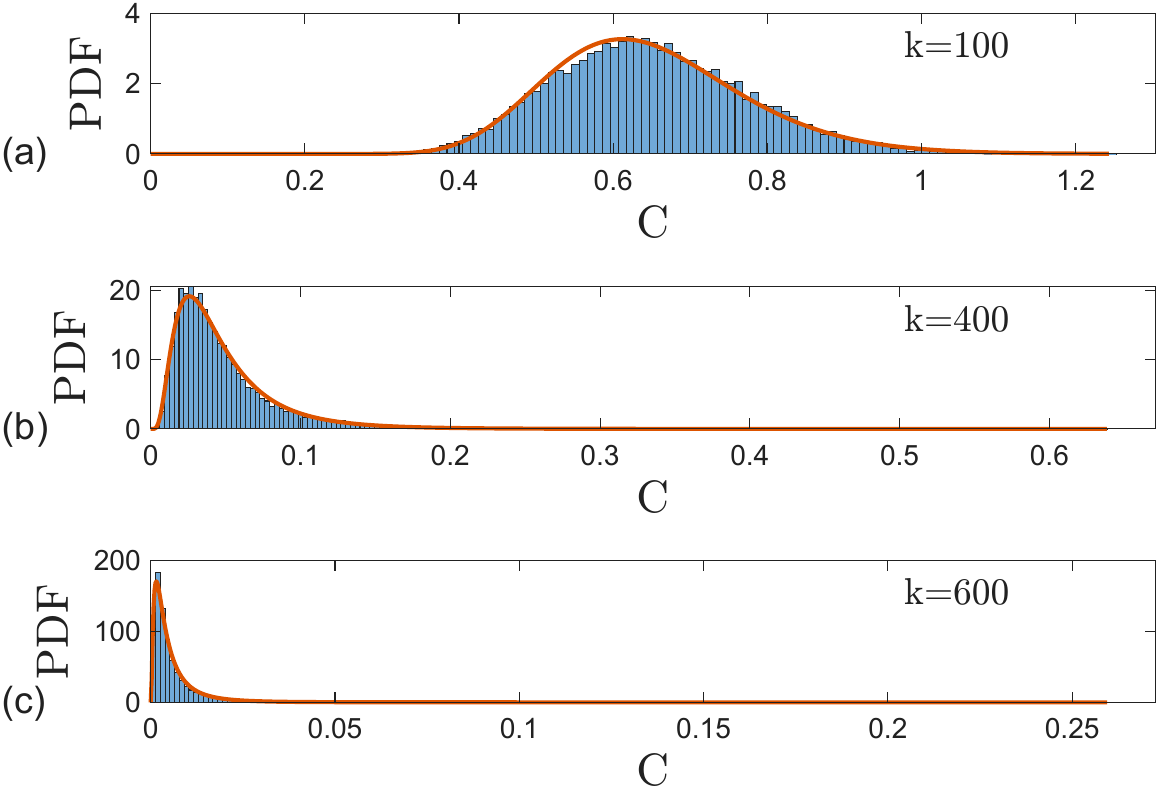}
    \caption{ 
        Histograms of the distribution of values of $C_k$ at different sites $k$ down the chain. The parameters are the same as those for \reffig{CorrelationsNearCritical}. The horizontal scales are chosen to include all values of $C$ in the random samples. When $C$ is small, there are some rare ``lucky'' samples with much larger values than most. 
                }
    \label{fig:DistributionsNearCritical}
\end{figure}

The first observation to be made about the behavior shown in \reffig{CorrelationsNearCritical} is that over most of the region shown the arithmetic mean and geometric means  are nearly indistinguishable from each other. However, a second observation is that down the chain, when the value of the correlation function is quite small, the two differ significantly. The inset shows the configuration-averaged correlations $\overline{C}$ and $C^{typ}$ on a logarithmic scale in the region around $k=1100$. By that point,  the arithmetic mean is larger than the geometric mean by as much as two orders of magnitude. The reason for this is made apparent in \reffig{DistributionsNearCritical} which shows a histogram of the corresponding values of the correlation function $C_k$ at three points along the chain: $k=100, 400$, and $600$. The range of the horizontal axis in each case is chosen so that all the values from the 20,000 samples are included. The red curves are fits to a log-normal probability distribution. As one moves down the chain, a few rare samples yield values of $C$ that are much larger than those of the vast majority of other samples. As a result, the arithmetic mean is dominated by these rare cases. 

It is perhaps tempting to attribute this to the presence of Griffiths phases, but there is reason for caution. The Appendix shows that a very similar phenomenon is evident in  the much simpler system of a disordered tight-binding chain (see Figures \ref{fig:GeomeanAnderson} and \ref{fig:DistributionAnderson}), a system without multiple phases. This suggests that the divergence between the two means may    result simply  from  the multiplicative  character of scattering from disorder in both the tight-binding and Ising chains.  We also see a similar separation between geometric and arithmetic means of $C$ for disordered Ising systems composed entirely of paramagnetic or ferromagnetic couplings. 

\begin{figure}[!tbp]
    \centering
    \includegraphics[width=\figwidth]{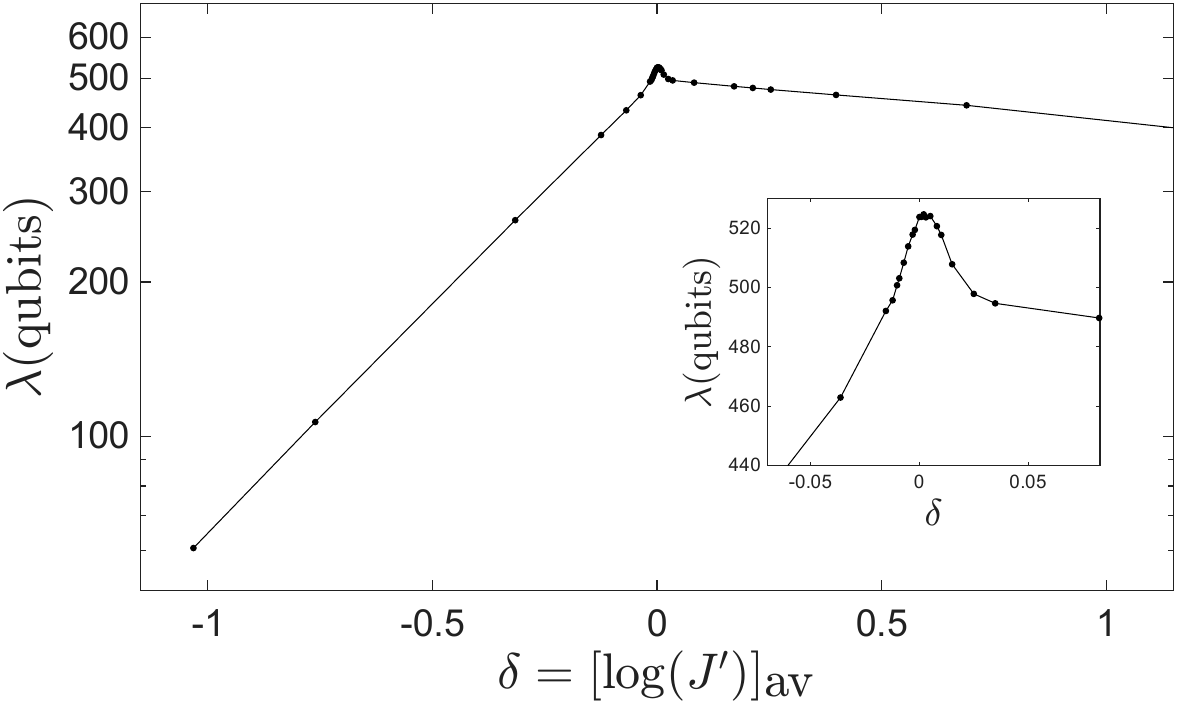}
    \caption{
        The correlation length for the disordered TFIM for $\Delta J'=0.6$ and various values of $J'_0\in [0.4, 3.3]$ corresponding to the values shown for $\delta$. The inset shows the peak on a linear vertical scale.
        }
    \label{fig:DecayLengthCritLogJ}
\end{figure}

Evidence for Griffiths phases is perhaps more apparent if we examine  correlations  in the region right around the quantum phase transition. \reffig{DecayLengthCritLogJ} shows the variation of  the correlation decay length $\lambda$ as the central coupling $J'_0$ (or equivalently $\delta$) is varied. Here $\Delta J'=0.6$, the threshold value $C_{\text{thresh}}=0.01$, and the number of samples $N_c$ varies from 10,000 to 30,000 for each value of $J'_0$ (more samples are required to achieve convergence near the transition). The time $t=160\tau$ is sufficiently long to show stationary behavior. In the paramagnetic domain to the left, $\lambda \sim (J'_0)^2$. In the ferromagnetic regime to the right, the correlation decay length varies quite slowly with $J'_0$. In a narrow region around $\delta=0$, the decay length is enhanced. The inset shows the peak with a linear vertical scale for $\lambda$. 

Using the language of quasiparticles,\cite{CalabreseCardy2006, Calabrese2007, RiegerIgloi2011} we also note that the peak in the correlation length at the critical point may be associated with the zero 
in the  quasiparticle excitation energy at the phase boundary seen in \reffig{EfastJp}. The disorder shifts the critical point from $J'=1$ to $\delta=0$ and changes a singularity to merely a peak.


\section{Connection to Su-Schrieffer–Heeger model \label{sec:SSHcomparison}}
The Appendix shows the qualitative similarity between the time evolution of the Lieb-Robinson correlation function $C_k(t)$ for the TFIM and the probability $P_R(k, t)$ (see \refeq{PRdef}) for a standard tight-binding chain. This similarity is quite general, including the decay due to random coupling disorder,  and extends to the statistics of  $C_k$  and $P_R$.  Because both quantities depend on the multiplicative effects of reflection and transmission at inhomogeneities, they both yield  log-normal distributions. 

We now show that in fact the expression for the Lieb-Robinson correlation function $C_k$ in  \refeq{C_eval_inh} can be rewritten in terms of the time evolution of the probability $P_R$ for a very specific single-particle tight-binding type Hamiltonian and a particular initial state. This makes the qualitative comparison in the Appendix quantitative and exact. 

The Su-Schrieffer–Heeger (SSH) model \cite{Su1979, Batra2020} describes the motion of a single particle on a 1D chain of $2N$ sites under the Hamiltonian

\begin{eqnarray}
H_{\mathrm{SSH}} &=  v\sum\limits_{k}^{N}  
\ket{{2k}}\bra{{2k-1}} +\text{h.c.} \nonumber \\
 & +   \sum\limits_k^{N-1} w_k  \ket{{2k+1}}\bra{{2k}} +\text{h.c.} .
\label{eq:SSHstandard}
\end{eqnarray}

Here the $2N$ sites are coupled together in $N$ unit cells and \( v \) and \( w_n \) are the intra-cell and inter-cell hopping amplitudes, respectively.
When all the $w_n$'s are the same, this model exhibits two  distinct phases depending on whether \( |v| < |w| \) or \( |v| > |w| \). 
We focus on the case when the inter-cell hopping terms $w_n$ are not necessarily the same, introducing possible disorder. 
The matrix representation of $H$ is given by
\begin{equation}
\bm{H}_{\text{\tiny SSH}} = 
\begin{pmatrix}
0 & v & 0 & 0 & 0 & 0 & \cdots & 0 \\
v & 0 & w_1 & 0 & 0 & 0 & \cdots & 0 \\
0 & w_1 & 0 & v & 0 & 0 & \cdots & 0 \\
0 & 0 & v & 0 & w_2 & 0 & \cdots & 0 \\
0 & 0 & 0 & w_2 & 0 & v & \cdots & 0 \\
\vdots & \vdots & \vdots & \vdots & \ddots & \ddots & \ddots & \vdots \\
0 & 0 & 0 & 0 & \cdots & w_{N-1} & 0 & v \\
0 & 0 & 0 & 0 & \cdots & 0 & v & 0
\end{pmatrix},
\label{eq:SSHmatrix}
\end{equation}
This bears a clear similarity to the matrix  $\bm{A}$ in \refeq{A_inh}.
We define
\begin{equation}
    \bm{H}_A\equiv -iA.
\end{equation}

\noindent The local gauge-transformation 
\begin{equation}
[\bm{U}]_{n,n'} = e^{i\pi n/2} \delta_{n,n'}, 
\end{equation}
changes the sign of the lower-triagonal elements and transforms $\bm{H}_A$ into the dimensionless SSH Hamiltonian,
\begin{equation}
\bm{H}'_{\text{\tiny SSH}}= \bm{U} \bm{H}_A \bm{U}^\dagger,
\end{equation}
if we make the identification  $w_n/v\leftrightarrow J'_n$. 

The time evolution of a state function on the SSH chain is  given by
\begin{equation}
\psi_{\text{\tiny SSH}}(t)=e^{-i 2\pi H'_{\text{\tiny SSH}} (t/\tau)}\psi_{\text{\tiny SSH}}(0). 
\end{equation}

\noindent We can therefore directly connect the time-evolution of the TFIM Lieb-Robinson correlation function $C_k(t)$ with the time evolution of the wavefunction on an SSH chain using the techniques shown in the Appendix for a standard tight-binding chain. 
\noindent We choose an initial state at the first SSH site
\begin{equation}
\braket{k|\psi_{\text{\tiny SSH}}(0)}=\delta_{k,1},
\label{eq:SSHinit}
\end{equation}

\noindent and calculate 

\begin{equation}
P^{\text{\tiny SSH}}(n,t)=
\left| \braket{n | e^{-i 2\pi H'_{\text{\tiny SSH}} (t/\tau)}\psi_{\text{\tiny SSH}}(0)} \right|^2,
\end{equation}

\noindent the probability of finding the particle on site $n$ at time $t$, given that it was on site $1$ at time $t=0$. If we define $P_R^{\text{\tiny SSH}}(k,t)$ for $k\in[1,N]$  to be the probability of finding the particle at time $t$ to the right of SSH site $2k-1$, 

\begin{equation}
P_R^{\text{\tiny SSH}}(k,t)=\sum_{m=2k}^{2N_q} P^{\text{\tiny SSH}}(m,t),
\end{equation}

\noindent the expression in \refeq{C_eval_inh} can  be written

\begin{equation}
C(k,t)=2\sqrt{P_R^{\text{\tiny SSH}}(k,t)}.
\label{eq:CktoSSH}
\end{equation}

\refeq{CktoSSH} links the spread of probability in an SSH chain to the spread of the Lieb-Robinson correlation function in a TFIM chain. Mathematically and in terms of computational ease, \refeq{CktoSSH} is  precisely equivalent to \refeq{C_eval_inh}. 
The basis states for $H'_{SSH}$ map onto the operators determined by the OPW method as illustrated in \reffig{pauli_walk_graph}, which it should be noted are the Majorana operators for the spin chain. The state-independent quantity $C_k(t)$ for the many-body TFIM for $N_q$ spins is connected by \refeq{CktoSSH} exactly to the evolution of the single-particle state defined on the $2N_q$ SSH sites with the initial state given by \refeq{SSHinit}.

\section{Discussion \label{sec:Discussion}}

Seeing the precise form that the Lieb--Robinson correlation function takes as quantum correlations move down a long Ising chain can be very revealing. This calculation is crucially enabled by the operator Pauli walk method \cite{Lent2024}. Figures \ref{fig:ferromagnetic_decay_plot} and \ref{fig:ferromagneticBoundcheck} show calculations on a 1000-qubit chain. To represent either a  quantum state or  quantum operators on the chain would require a basis of size $2^{1000} \approx 10^{301}$, a number that greatly exceeds the number of particles in the visible universe. By contrast, evaluating \refeq{C_eval_inh} for $C_k(t)$ requires a basis size of only 2000. This is all the more important for assessing directly the effects of disorder because of the need to average over many random configurations. Of course, the quantum state contains vastly more information than the correlation function. 

As we have seen, another feature of the OPW result in \refeq{C_eval_inh} beyond treating arrays that are very long, is that it can be used to compute the LR correlation function at any time, including very long times, without time-marching or Trotterization. This makes it possible to see directly that the light-cones for the disordered Ising chain are not logarithmic, but in fact become vertical (zero-velocity), in agreement with the bound of Hamza, Sims, and Stolz \cite{Hamza2012, Stolz2017}.

The analysis of the single-particle tight-binding chain in the Appendix A illustrates the key features of disorder-induced localization using the same approach we subsequently employed in Section \ref{sec:DisorderedIsing}. The examination of the statistics of $P_R(x)$ also shows that one feature observed in the Ising results---the difference between the arithmetic and geometric means---is actually a generic feature of the multiplicative nature of multiple-scattering systems \cite{RiegerIgloi2011}.

The method we have employed is not general-purpose.  The closed nature of the walk in the OPW method is a consequence of the fact that the Ising Hamiltonian can be written as a  quadratic in the Majorana operators which form the operator basis set (for example, those in \refeq{4q_sigma_list}). For example, including next-near-neighbor interactions would yield an explosion in the number of operators required. The method is also limited to correlations between the first qubit in the chain and the $k^{\text{th}}$ qubit.

In reference \cite{Igloi2017}, Ro\"{o}sz, Lin, and Igl\'{o}i use a Majorana basis for calculating the relaxation of the average magnetization from a quench for the disordered TFIM. We show in Appendix B  that using such a basis provides an alternate derivation for the results from our operator Pauli walk method.

The mapping of the Ising  Lieb-Robinson correlation function, which is importantly state-independent, onto the evolution of a \emph{particular} state for the SSH model provides another interpretation of the main result \refeq{C_eval_inh} and may suggest further avenues for exploration. The linkage between the two models is well known, as is the occurrence of zero-energy edge modes in both.

It would be a mistake to  conclude that the many-body localization seen in the TFIM is just the Anderson localization of a  quasiparticle. The full excitation spectrum of the TFIM Hamiltonian includes the presence of all possible quasiparticles. We have not limited the analysis to the transport of a single quasiparticle; all are included in the correlation function. Similarly, there is no restriction to only low-energy excitations. The localization in this many-body system must be localization of all quasiparticle excitations. 

The primary result of this work is the calculation  of the Lieb-Robinson correlation function, rather than a bound on it, in the case of disorder.   Quantum correlations and information are localized and not able to propagate farther down the chain. Extending the techniques that made this possible is a subject for further study.

\appendix

\section{The tight-binding chain: 
Disorder and Anderson localization \label{sec:TBchain} }

\subsection{Basic model}

We first consider a single particle moving on a 1D chain with no disorder. The Hamiltonian for a tight-binding chain of $N$ sites can be written
\begin{equation}
\hat{H} = - \gamma  
\sum_{k=1}^{N-1}  \Bigl[ \ket{k+1}\bra{k} + \ket{k}\bra{k+1}   \Bigr]
\label{eq:Htb0}
\end{equation}
where  the hopping matrix element is $\gamma$, which we take to be positive. The on-site potentials are taken to be zero because they play no role here; this simply sets the zero of energy. The characteristic time is given by
\begin{equation}
\tau \equiv \frac{\pi \hbar}{\gamma}.
\end{equation}
In the thermodynamic
limit, the energy eigenstates can be labeled by a wavenumber $q$ and the eigenenergies are given by

\begin{equation}
E(q) = - 2 \gamma \cos(q).
\end{equation}
We define the velocity $v$ as the maximum group velocity, thus
\begin{equation}
v\, \tau = \tau\frac{1}{\hbar} 
\left . \frac{\partial E} {\partial q} \right|_{max}
= 2 \pi.
\label{eq:Vtb}
\end{equation}

\subsection{Disorder-free case}

We are interested in the case in which the particle is initially localized on the first site so $\Psi(k,0) = \delta_{k,1}$. We solve for the time-dependent wavefunction by directly performing the unitary transformation
\begin{equation}
\Psi(k, t) = e^{-i \hat{H} t/\hbar} \ \Psi(k,0).
\end{equation}
The probability density at site $k$ and time $t$ is then $P(k,t) =\left| \Psi(k,t) \right|^2.$

\begin{figure}[!tbh]
    \centering
    \includegraphics[width=\figwidth]{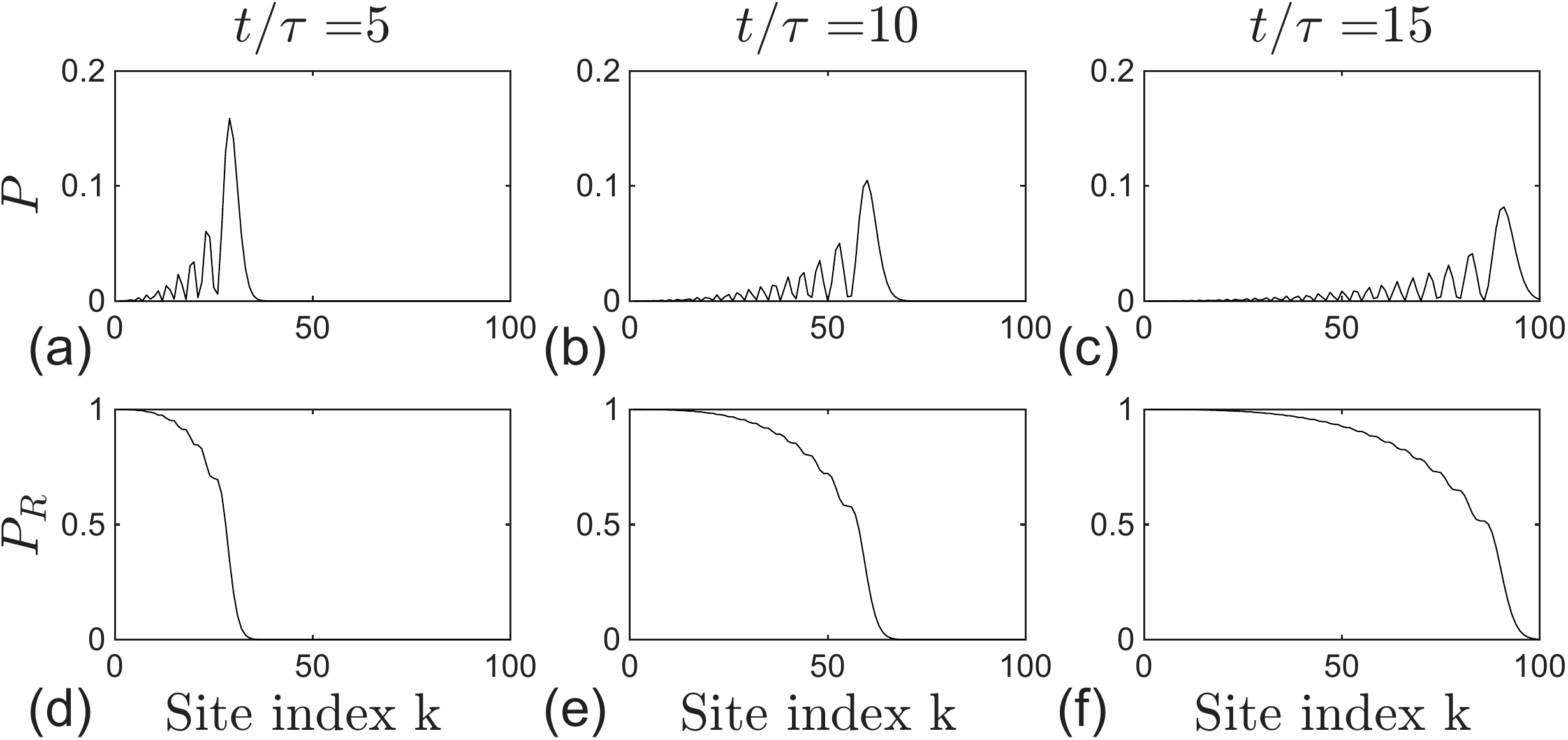}
    \caption{Snapshots of the probability for a tight-binding chain of 600 sites. $P(k,t)$, the probability of finding the particle at site $k$, is shown at different times in (a)-(c). The probability $P_R(k,t)$ of finding the particle to the right of site $k$ is shown in (d)-(f).
    }
    \label{fig:TBsnapshotDgamma0}
\end{figure}

Figure \ref{fig:TBsnapshotDgamma0}(a)-(c) shows snapshots of the probability density for a line of $N=600$ sites at times $t/\tau = [5,10,15]$. The probability peak moves down the chain with the velocity given by \refeq{Vtb} (i.e., $v= 2\pi$ sites/$\tau$).

To connect the evolution of the probability $P(k,t)$ with the Lieb-Robinson correlation function in the TFIM, 
we first define 

\begin{equation}
P_R(k,t) = \sum_{k'=k+1}^{N} P(k',t),
\label{eq:PRdef}
\end{equation}
the probability of finding the particle somewhere to the right of site $k$ at time $t$.  Figure \ref{fig:TBsnapshotDgamma0}(d)-(f) shows snapshots of $P_R$ for the same times as above. The front moves down the chain with velocity $v$.  The sites ahead of the front do not yet ``know'' about the particle that was at site 1 at time $t=0$. 

The light-cone plot of $P_R(k,t)$ for this system is shown in Figure \ref{fig:TBLightConePlots}(a). The dashed red line corresponds to the front motion. It has a slope of $1/(2\pi)$ corresponding to the velocity in \refeq{Vtb}; because the time is shown on the vertical axis, the front velocity is the inverse of the line's slope.

\subsection{Disordered case--Anderson localization}
We now consider the case where the hopping matrix elements between sites vary randomly down the chain. We are not considering the case with varying on-site potentials simply because variable hopping is most clearly analogous to the TFIM with disordered couplings between spins, which is our primary focus. 
The Hamiltonian is then
\begin{equation}
\hat{H} = -  
\sum_{k=1}^{N-1}  \gamma_k \Bigr[ \ket{k+1}\bra{k} + \ket{k}\bra{k+1}   \Bigl].
\label{eq:HtbDisorder}
\end{equation}

The matrix elements $\gamma_k$ are independently drawn from a uniform probability distribution centered at $\gamma_0$ with width $\Delta\gamma$:

\begin{equation}
P(\gamma)=
\begin{cases}
\dfrac{1}{\Delta \gamma} & \text{for } \gamma \in 
\left[\gamma_0 - \tfrac{\Delta \gamma}{2},\, \gamma_0 + \tfrac{\Delta \gamma}{2}\right],\\[4pt]
0 & \text{otherwise}.
\end{cases}
\label{eq:UniformPDFgamma}
\end{equation}

\begin{figure}[!tbp]
    \centering
    \includegraphics[width=\figwidth]{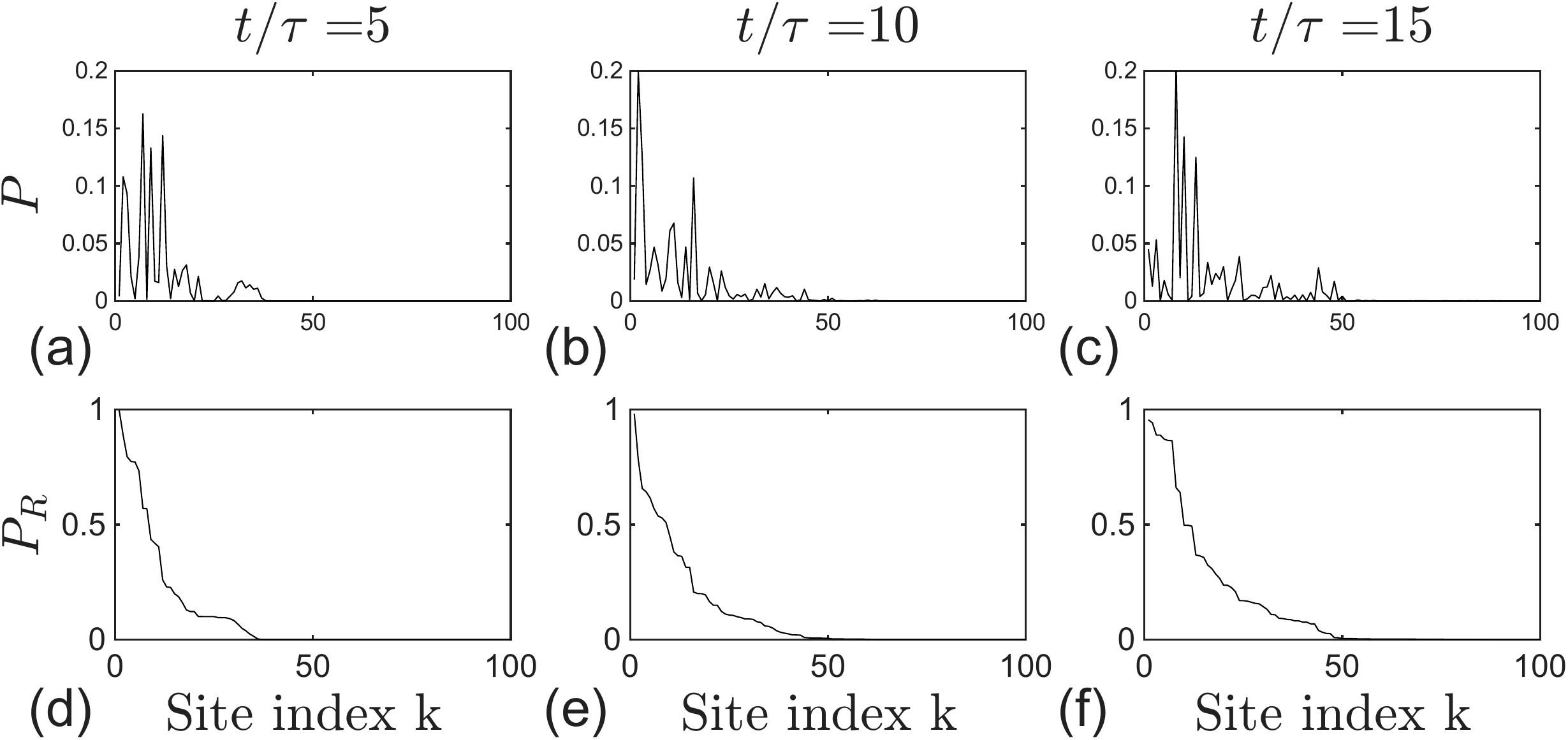}
    \caption{Snapshots of the probability for the tight-binding chain of 600 sites with  disorder.  The site probability $P(k,t)$ is shown in (a)-(c). The probability of finding the particle to the right of site $k$, $P_R(k,t)$ is shown in (d)-(f). The disorder in the hopping matrix element is characterized by $\Delta\gamma/\gamma_0=1$.  
    }
    \label{fig:TBsnapshotDgamma1}
\end{figure}

\FloatBarrier

\begin{figure}[!tbp]
  \begin{minipage}{0.85\columnwidth}
    \makebox[16pt][l]{\textbf{(a)}\hspace{1em}}%
    \includegraphics[width=\linewidth]{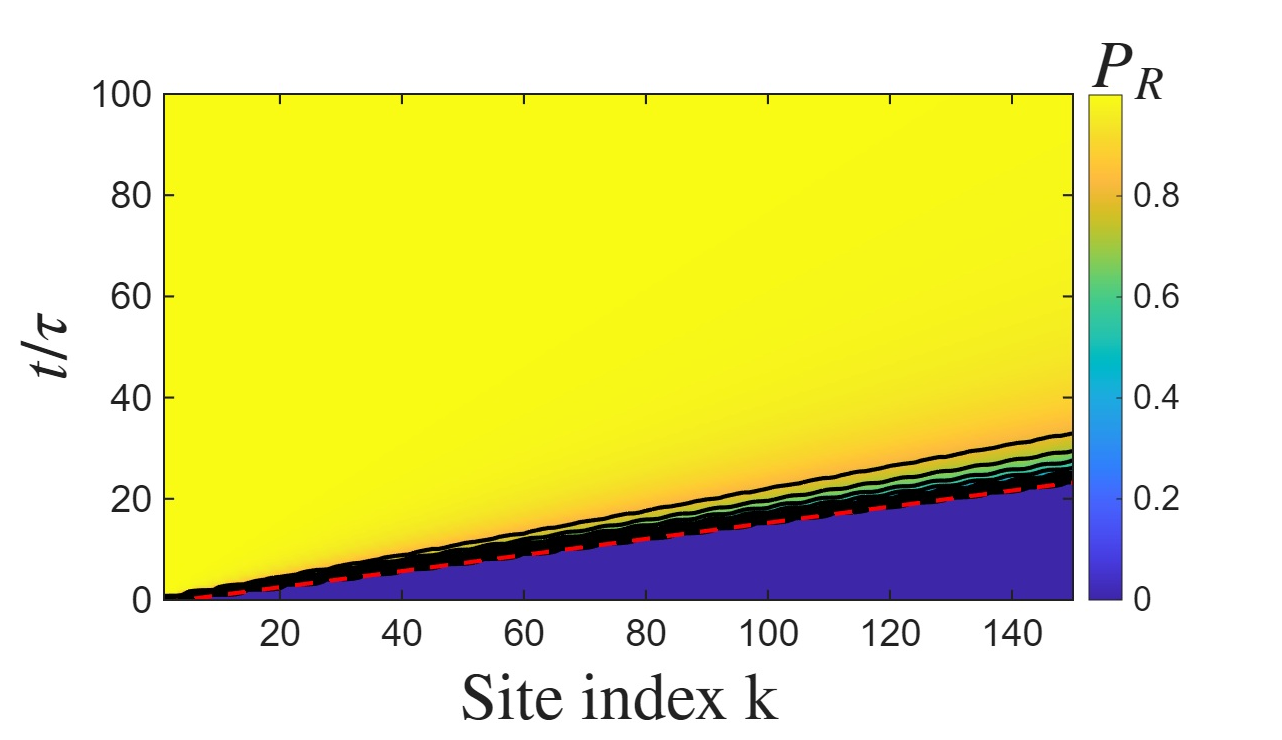}
  \end{minipage}

  \vspace{1em}

  \begin{minipage}{0.85\columnwidth}
    \makebox[16pt][l]{\textbf{(b)}\hspace{1em}}%
    \includegraphics[width=\linewidth]{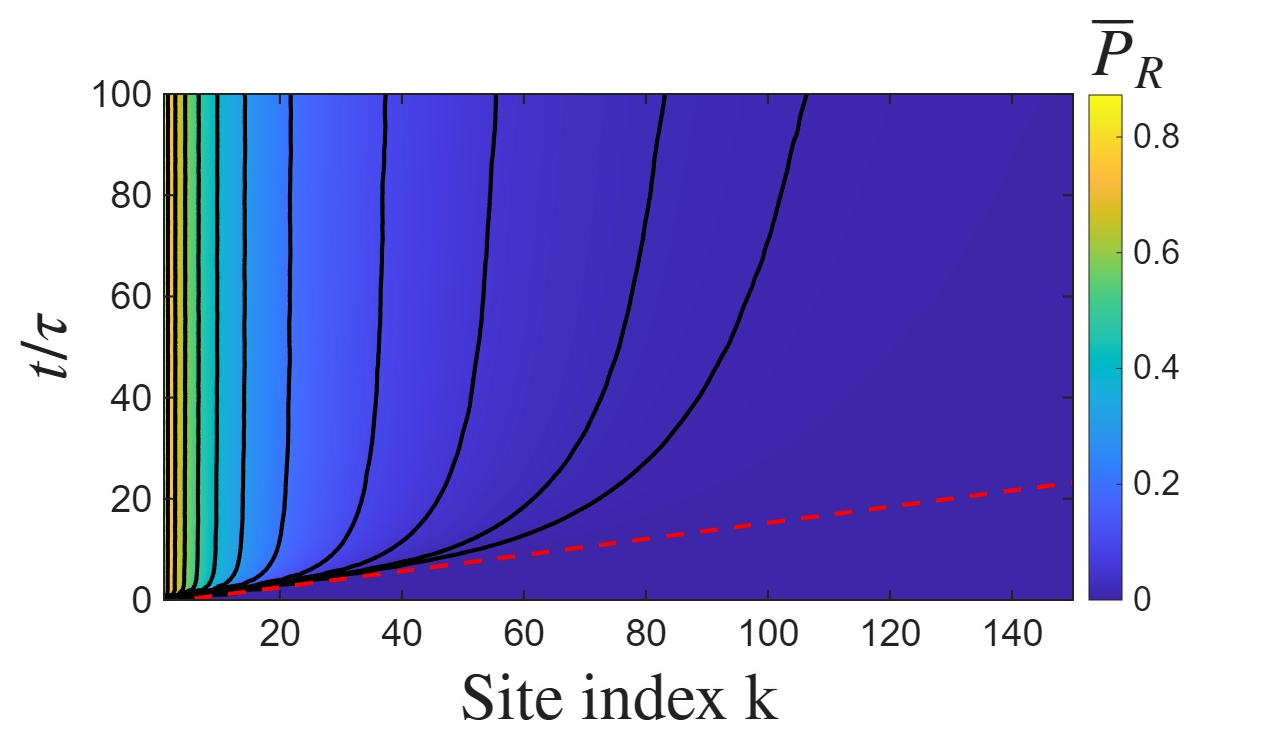}
  \end{minipage}

  \begin{minipage}{0.85\columnwidth}
    \makebox[16pt][l]{\textbf{(c)}\hspace{1em}}%
    \includegraphics[width=\linewidth]{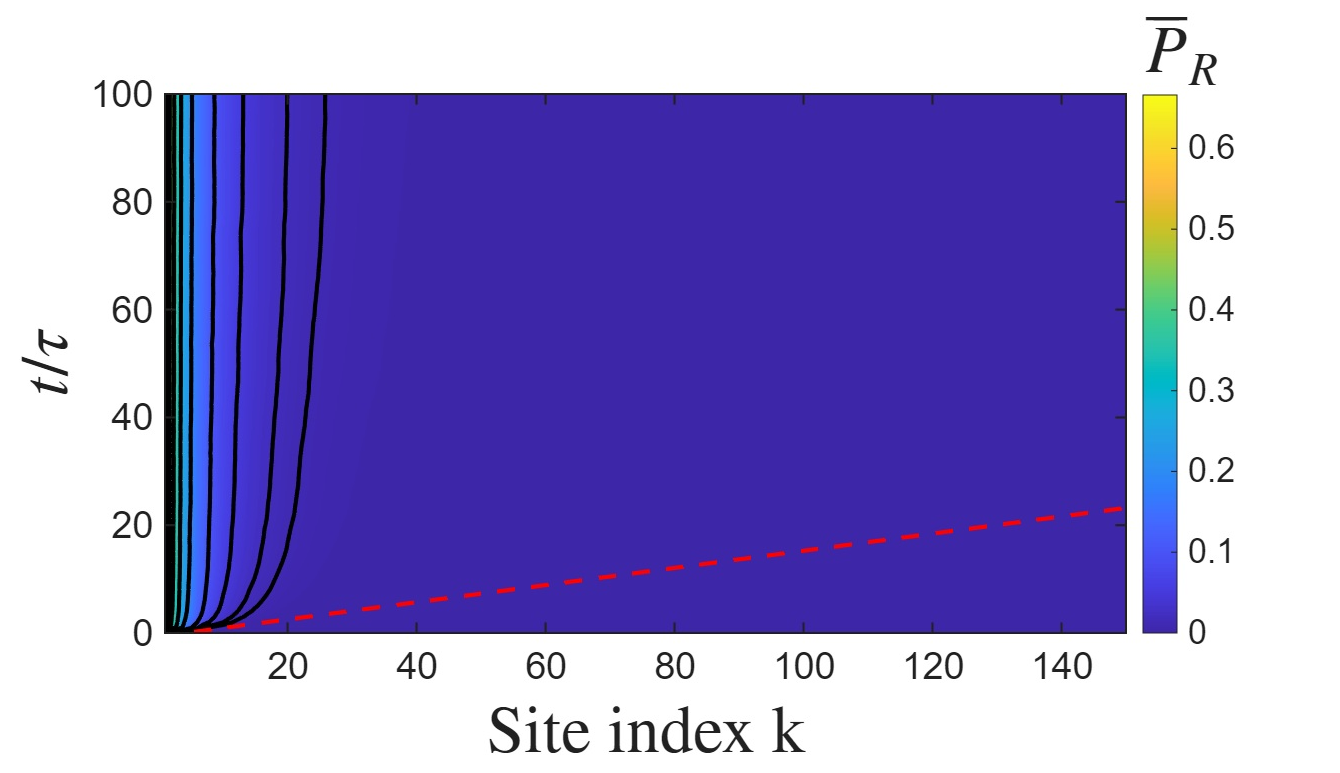}
  \end{minipage}

  \caption{ 
  Light-cone plots for 1D tight-binding chain of 300 sites.  (a) The color indicates the probability of the particle being found to the right of site $k$, $P_R(k,t)$ for the case of no disorder, $\Delta\gamma=0$.  The color indicates the disorder-averaged probability $\overline{P}_R(k,t)$ for   (b) weak disorder with $\Delta\gamma/\gamma=1$,   and (c) strong disorder with $\Delta\gamma/\gamma=2$. The red dashed line in each plot corresponds to the velocity of \refeq{Vtb}. }
  \label{fig:TBLightConePlots}
\end{figure}

A particular set of $N-1$ randomly chosen $\gamma_k$'s constitutes a specific disorder configuration for the tight-binding chain. We solve for the unitary time evolution of the state function under (\ref{eq:HtbDisorder}) and compute the probabilities $P$ and $P_R$ as before. Figure \ref{fig:TBsnapshotDgamma1} shows snapshots of the time evolution of both probabilities for a specific disorder configuration chosen   from a distribution with $\Delta\gamma/\gamma_0=1$. The snapshots are at times  $t/\tau = [5,10,15]$, where $\tau\equiv \pi\hbar/\gamma_0$. It is clear even in this single sample that, when compared with  the non-disordered case of Figure \ref{fig:TBLightConePlots}(a), the particle propagation is impeded by the multiple reflections caused by the disorder.

Of more interest than the behavior of a specific disorder configuration is the average behavior. We define $\overline{P}_R$ as the configuration average of $P_R$,
\begin{equation}
    \overline{P}_R(k,t)= \frac{1}{N_c} \sum_{m_c}^{N_c} P^{(m_c)}_R(k,t), 
    \label{eq:TBPR}
\end{equation}
where $m_c$ is the index of a specific disorder configuration and $N_c$ is the total number of such configurations sampled. 


\FloatBarrier

Light-cone plots of $\overline{P}_R$ for the cases of $\Delta\gamma/\gamma_0=1$ (weak disorder) and $\Delta\gamma/\gamma_0=2$ (strong disorder) are shown in
Figures \ref{fig:TBLightConePlots}(b) and \ref{fig:TBLightConePlots}(c) respectively. Here $N=300$ sites, though the plots only extend over half that length. The number of disorder configurations in the average is  $N_c$=20,000. Contours of constant $\overline{P}_R=[0.01, 0.02, 0.05, 0.1, 0.2, 0.3, 0.4, 0.5, 0.6, 0.7, 0.8]$ are indicated by black lines. The red dashed line in each figure corresponds to ballistic propagation with velocity $v=2\pi/\tau$ as seen in Figure \ref{fig:TBLightConePlots}(a).

The light-cone plots of Figures \ref{fig:TBLightConePlots}(b) and \ref{fig:TBLightConePlots}(c) reveal that the disorder suppresses particle motion. This is the well-established single-particle Anderson localization in 1D. The light-cone curves (isocontours of $\overline{P}_R$)  are bent upward by the presence of the disorder. Their shape is neither linear nor logarithmic. As time goes on, they in fact become vertical, indicating no further motion down the chain---zero propagation velocity. 

\begin{figure}[!tbp]
    \centering
    \includegraphics[width=\figwidth]{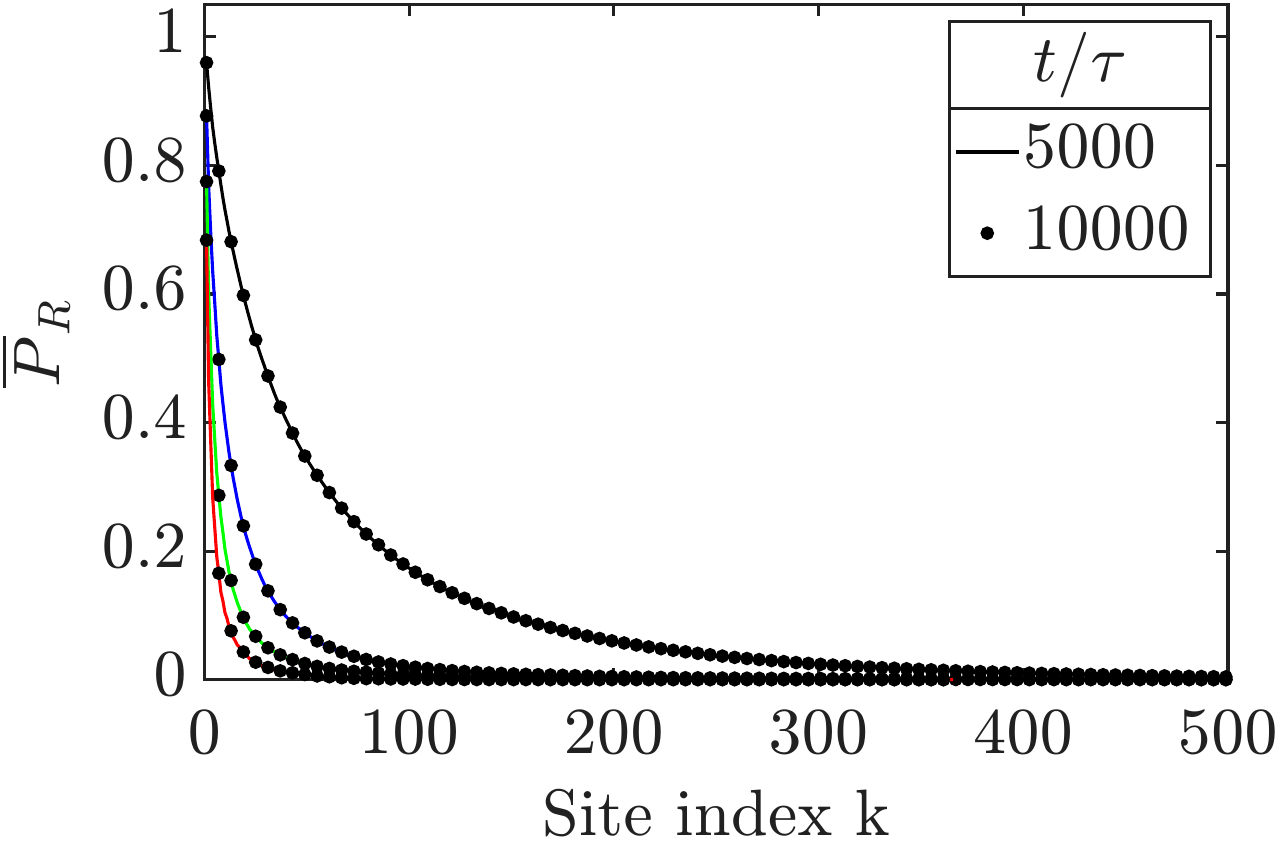}
    \caption{Configuration averaged probability $\overline{P}_R$ for two long times. The average is over 2000 sample configurations and $N=600$. The disorder is characterized by $\Delta\gamma/\gamma_0=0.5$ (black), 1 (blue), 1.5 (green) and 2 (red).
    }
    \label{fig:TBLongDecayTimes}
\end{figure}

This can be seen more explicitly by considering $\overline{P}_R$ at very long times. Figure \ref{fig:TBLongDecayTimes} shows $\overline{P}_R$ for two long times, $t/\tau=$5,000 and 10,000, calculated for four different values of the probability distribution width. For each probability distribution, the result at the two times is identical, indicating that by these times there is no  particle transport  down the chain. 

\subsection{Average versus typical values}
We have defined the quantity $\overline{P}_R(k,t)$ in \refeq{TBPR} as the arithmetic mean over the set of sample disorder configurations. Variations in the hopping matrix element $\gamma_k$ can be viewed as constituting a series of tunneling barriers and wells whose transmissions combine multiplicatively. This suggests using a geometric  mean (log-average) to characterize the behavior of the set of random configurations. We define therefore the typical value of $P_R$ by

\begin{equation}
    P^{typ}_R(k,t)= \exp\left( \frac{1}{N_c} \sum_{m_c}^{N_c} \log\left(P^{(m_c)}_R(k,t)\right) \right)
    \label{eq:PRtypAnderson}
\end{equation}

\begin{figure}[bt]
    \centering
    \includegraphics[width=\figwidth]{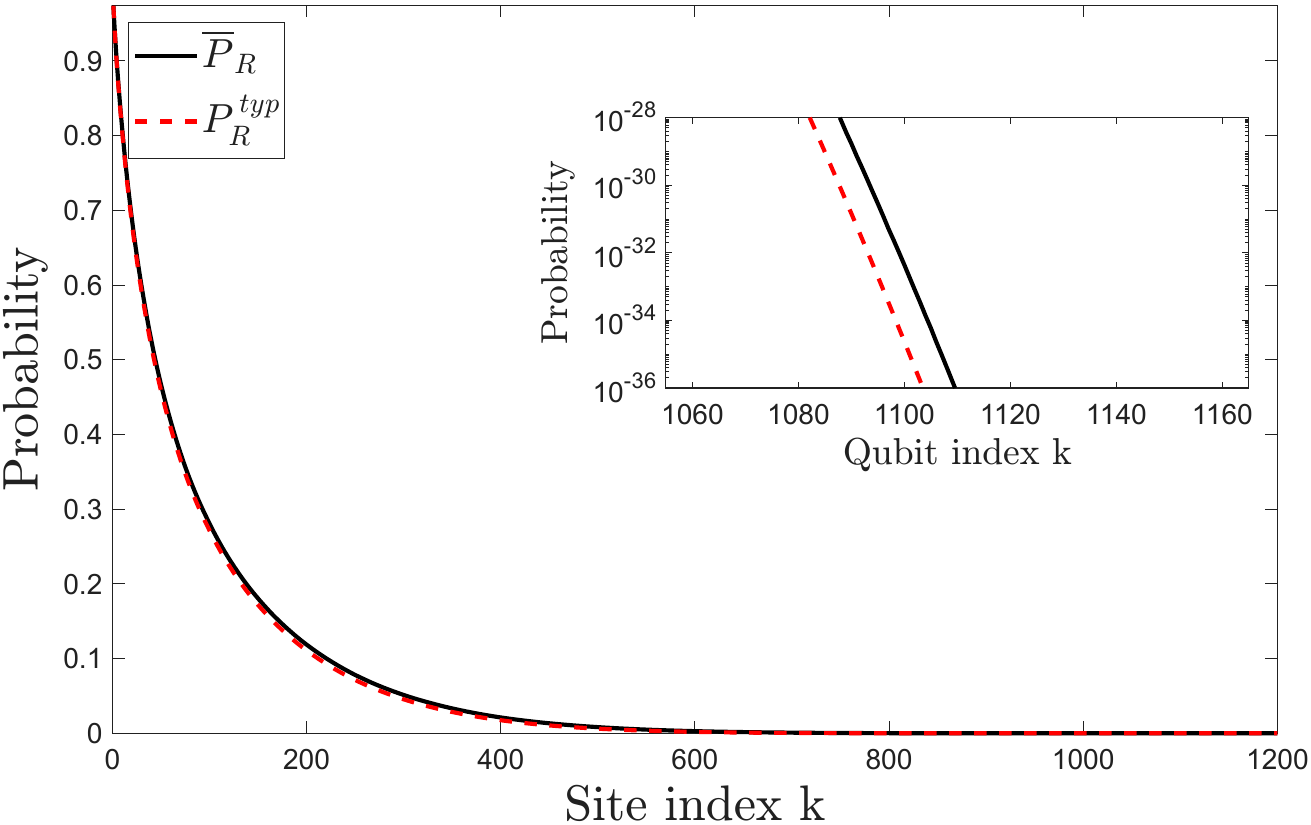}
    \caption{The arithmetic mean and typical values of $P_R$ over a set of $N_c=10,000$ samples. Here the tight-binding chain has inter-site coupling $\Delta \gamma/\gamma_0 =0.4$ with $N=1200$ sites and is evaluated at time $t/\tau=160$. The inset shows the region around $k=1100$ on a log-scale.
    }
    \label{fig:GeomeanAnderson}
\end{figure}

Figure  \ref{fig:GeomeanAnderson} compares $\overline{P}_R$ and $P^{typ}_R$  for a disordered chain of $N=1200$ sites with $\Delta \gamma/\gamma =0.4$, $N_c=10,000$, and $t/\tau=160$. By this time, multiple paths of interference have essentially extinguished propagation down the chain. On the left, where the average $P_R$ is large, the two means differ only very slightly. For the light-cone diagrams of Figure \ref{fig:TBLightConePlots},  the two measures would be indistinguishable from each other.  But near the right, when the probability is very small, the distinction is worth noting. The inset of Figure  \ref{fig:GeomeanAnderson} shows the two means on a logarithmic scale in this region farther down the chain. The value of $\overline{P}_R$ in this region can be two orders of magnitude larger than  $P^{typ}_R$. 

\begin{figure}[bt]
    \centering
    \includegraphics[width=\figwidth]{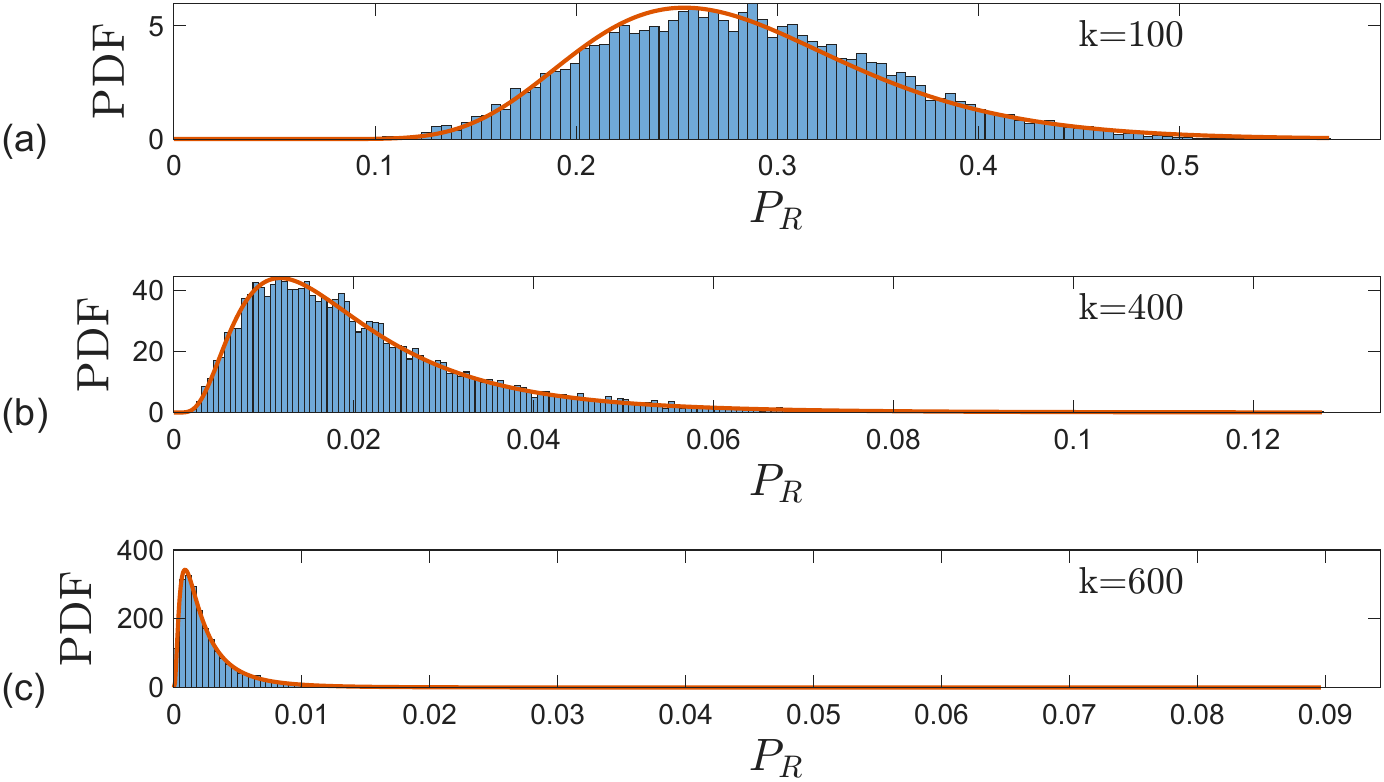}
    \caption{The probability distribution function of the values of $P_R$ for the same chain as in Figure \ref{fig:GeomeanAnderson}.
    Distributions are shown with a log-normal fit at three values of site index $k$.
    }
    \label{fig:DistributionAnderson}
\end{figure}

The reason for this is made clear by considering the probability distribution function (PDF) of $P_R$ in the ensemble of samples. Figure \ref{fig:DistributionAnderson} shows the probability distribution at three different points in the chain ($k=100$, $k=400$, and $k=600$). The red line shows a fit to a log-normal distribution, which has a characteristically long tail. The figure's horizontal axis is chosen to include all the values of $P_R$ which occur in the $10,000$ samples. One can see in Figure \ref{fig:DistributionAnderson}(c) that some samples have a value of $P_R$ that is more than an order of magnitude larger than the  peak of the distribution. The arithmetic mean $\overline{P}_R$ is dominated by these relatively few ``lucky'' samples with a larger probability  extending farther down the chain. The typical value defined by \refeq{PRtypAnderson} is then more useful.


\section{Formulation with the Majorana basis \label{sec:MajoranaBasis} }

\subsection{Majorana definition and properties \label{sec:MajDefs}}

There are $2N_q$ Majorana operators $\gamma_j$, one with an odd index and one with an even index for  each qubit $k$. They are defined by
\begin{equation}
\begin{aligned}
P_k&\equiv \prod_{j<k} X_j, \quad\quad P_1=I\\
\gamma_{2k-1}&\equiv P_k Z_k,\quad\quad\gamma_{2k}\equiv P_kY_k.
\end{aligned}
\end{equation}
Each operates on the Hilbert space of the full  chain, so each can be represented as a matrix of size $2^{N_q} \times 2^{N_q}$.
Note that these $\gamma_j $ are precisely the operators  used in the Pauli walk described in \ref{sec:PauliWalk}, as illustrated in  \refeq{4q_sigma_list}.

The Majoranas satisfy 
\begin{align}
\gamma_{k}^2&=I, \quad\quad \gamma_k=\gamma_k^{\dagger},\\
\{\gamma_i, \gamma_j\} &= 2\delta_{i,j}I, \\
[\gamma_m, \gamma_n] &=  2 \gamma_m \gamma_n, \quad m\neq n,\\
[\gamma_j \gamma_k, \gamma_{\ell} ] &= 2\delta_{\ell,k}\gamma_j - 2\delta_{\ell, j}\gamma_k.
\label{eq:commutatorGammaGamma}
\end{align}

Pauli operators can be expressed in terms of the Majorana basis,
\begin{align}
X_k &= \ii\, \gamma_{2k-1} \gamma_{2k}\\
Z_k &= \ii^{k-1} \gamma_1 \gamma_2 \gamma_3 \cdots \gamma_{2k-1} \label{eq:Zkprodgammas}\\
Z_kZ_{k+1}&=\ii\gamma_{2k}\gamma_{2k+1}\\
Z_1 &= \gamma_1, \quad \quad Y_1 = \gamma_2.
\end{align}

The full set of index-ordered Majorana strings, e.g., $G=\gamma_3\gamma_5\gamma_7$,  forms an orthonormal basis for the whole operator space with the normalized Frobenius inner product, 
\begin{equation}
\braket{\mathcal{O}_1,\mathcal{O}_2}= \frac{1}{2^{N_q}} 
{\rm{Tr}}(\mathcal{O}_1^{\dagger}\mathcal{O}_2),
\end{equation}
with the induced norm $\|\mathcal{O}\|=\sqrt{\braket{\mathcal{O},\mathcal{O}}}$, as in \refeq{FroebNorm}. Two Majorana strings which differ in any index are orthogonal. 


\subsection{Majorana time development for a quadratic Hamiltonian}
The Heisenberg equation of motion for $\gamma_\ell$ is
\begin{equation}
    \frac{d \gamma_\ell(t)}{dt}= \frac{\ii}{\hbar} [H,\gamma_\ell(t)]
    \label{eq:GammaEqOfMotion1}
\end{equation}

\noindent Consider a general Hamiltonian which is quadratic in Majorana operators. 

\begin{equation}
H=\frac{\ii}{4}\sum_{m,n=1}^{2N_q}\mathcal{A}_{mn}\gamma_m\gamma_n
\label{eq:HquadraticA}
\end{equation}
or equivalently
\begin{equation}
H=\frac{\ii}{4}\bm{\gamma}^{T}\mathcal{A}\bm{\gamma}
,
\qquad
\bm{\gamma}=
\begin{pmatrix}
\gamma_1\\
\gamma_2\\
\vdots\\
\gamma_{2N_q}
\end{pmatrix}.
\end{equation}
The matrix $\mathcal{A}$ is real and antisymmetric, so $H$ is Hermitian. The TFIM is a special case of this more general form. We will first find the Lieb-Robinson correlation function applicable to any quadratic Majorana Hamiltonian.

We can evaluate the commutator of a Majorana with the Hamiltonian using \refeq{commutatorGammaGamma},
\begin{align}
[H,\gamma_\ell]&= \frac{\ii}{4} 
\sum_{j,k=1}^{2N_q} \mathcal{A}_{jk}[\gamma_j \gamma_k,\gamma_\ell]\\
&= -\ii \sum_{j=1}^{2N_q} \mathcal{A}_{\ell j}\gamma_j.
\end{align}

\noindent Therefore \refeq{GammaEqOfMotion1} becomes

\begin{equation}
\frac{d \gamma_\ell}{dt}= \frac{1}{\hbar}
\sum_{j} \mathcal{A}_{\ell j} \gamma_j,
\end{equation}

\noindent or, using the vector of operators
\begin{equation}
\frac{d \bm{\gamma}}{dt}= \frac{1}{\hbar}  \mathcal{A} \bm{\gamma}.
\end{equation}
We can write the solution as
\begin{equation}
    \bm{\gamma}(t)=e^{(\mathcal{A}/\hbar)t} \bm{\gamma}(0).
\end{equation}

It is useful to define this time evolution matrix $R(t)$ as
\begin{equation}
    R(t)=e^{(\mathcal{A}/\hbar)t},
\end{equation}
so
\begin{equation}
    \bm{\gamma}(t)=R(t)\, \bm{\gamma}(0),
\end{equation}
or equivalently,
\begin{equation}
    \gamma_k(t)=\sum_m R_{km}(t)\,\gamma_m.
\end{equation}
$R$ is a $2N_q \times 2N_q$ real orthogonal matrix.

\subsection{Lieb-Robinson correlation function for a quadratic Hamiltonian}
The Lieb-Robinson operator is
\begin{equation}
\hat{C}_k(t)\equiv [Z_1(t), Z_k].
\end{equation}

\noindent To evaluate this we can use \refeq{Zkprodgammas} and
\begin{equation}
Z_1(t)= \gamma_1(t) =\sum_m R_{1m}(t)\gamma_m
\end{equation}
to obtain
\begin{equation}
\hat{C}_k(t)=\ii^{k-1}\sum_{m=1}^{2N_q} R_{1m}(t) [\gamma_m, \gamma_1 \gamma_2 \gamma_3 \cdots \gamma_{2k-1} ]
\label{eq:Chat11}
\end{equation}
For convenience, we define 
\begin{equation}
M_k\equiv \gamma_1 \gamma_2 \gamma_3 \cdots \gamma_{2k-1}.
\end{equation}
One can show that 
\begin{equation}
[\gamma_m,M_k]=
\begin{cases}
    0 & \text{if } m\le 2k-1 \\
    2\gamma_m M_k,  & \text{if } m\ge 2k. 
\end{cases}
\end{equation}
Therefore,  (\ref{eq:Chat11}) becomes
\begin{equation}
\hat{C}_k(t)=\ii^{k-1}\sum_{m=2k}^{2N_q} 2R_{1m}(t) \gamma_m M_k, 
\end{equation}
where the sum over $m$ now starts at $2k$.

The Lieb-Robinson correlation function is the normalized Frobenius norm of this operator. The Majorana strings $\gamma_m M_k$ are orthonormal, so we obtain simply

\begin{align}
C_k(t)&= 2\sqrt{\sum_{m=2k}^{2N_q} R_{1m}^2(t)}\\
&= 2\sqrt{ \sum_{m=2k}^{2N_q} \left( \left[ e^{(\mathcal{A}/\hbar)t} \right]_{1m} \right)^2
}
\label{eq:CkRsquared}
\end{align}

This is a general result for a quadratic Majorana Hamiltonian. It remains  to find $\mathcal{A}$  for the TFIM and relate this to our result in \refeq{C_eval_inh}.

\subsection{The TFIM in Majorana basis}

It can be shown from the fundamental relations in 
Section \ref{sec:MajDefs} that the TFIM Hamiltonian \refeq{H_dimful} can be written as
\begin{equation}
    H=-\ii  \sum_{k=1}^{N_q-1} J_k\gamma_{2k}\gamma_{2k+1} - \ii  \,\gamma\sum_{k=1}^{N_q} \gamma_{2k-1}\gamma_{2k},
\end{equation} 
where we use $\gamma$ here (possibly confusingly) for the energy of the transverse field term as in \refeq{H_dimful}.
This can be written in the form of \refeq{HquadraticA}
with 
\begin{equation}
\mathcal{A} \;=\;-2\gamma
\begin{pmatrix}
0      & 1     & 0      & 0      & 0       & 0      & \cdots   & 0  \\
-1     & 0     & J_1'   & 0      & 0       & 0      & \cdots   & 0  \\
0      & -J_1' & 0      & 1      & 0       & 0      & \cdots   & 0  \\
0      & 0     & -1     & 0      & J_2'    & 0      & \cdots   & 0  \\
0      & 0     & 0      & -J_2'  & 0       & \ddots & \vdots   & \vdots \\
0      & 0     & 0      & 0      & \ddots  & \ddots & J_{N_q-1}' & 0  \\
\vdots & \vdots& \vdots & \vdots & \cdots  & -J_{N_q-1}' & 0    & 1 \\
0      & 0     & 0      & 0      & \cdots  & 0      & -1       & 0 
\end{pmatrix}.
\label{eq:calA_inh}
\end{equation}

Comparing with \refeq{A_inh}, we see that
\begin{equation}
\mathcal{A}=-2\gamma A.
\end{equation}
and 
\begin{equation}
    R(t)=e^{\mathcal{A}t/\hbar}=e^{-2\gamma A t/\hbar}.
\end{equation}
Because $\tau=\pi\hbar/\gamma$, we can write this as
\begin{equation}
R(t)=e^{-2\pi(t/\tau)A},
\end{equation}
and the expression for $C_k(t)$ in \refeq{CkRsquared} becomes
\begin{equation}
C_k(t) \;=\;
2\sqrt{\sum_{m = 2k}^{2N_q}
\left|\, \left[
\exp\!\left(-2 \pi \tfrac{t}{\tau}\,A\right)
\right]_{1,m} \right|^2 },
\label{eq:C_eval_inhMaj}
\end{equation}
which is identical to the main operator Pauli walk result in \refeq{C_eval_inh}.



%

\end{document}